\documentclass[graphics, twocolumn, usenatbib]{mn2e}
\pdfoutput=1
\usepackage{times}
\usepackage{natbib} 
\usepackage{epsfig} 
\usepackage{graphicx} 
\usepackage{color}
\usepackage{aas_macros} 
\usepackage{amssymb}
\usepackage{amsmath}
\usepackage[title]{appendix}
\usepackage{hyperref}	
\hypersetup{colorlinks=true,linkcolor=blue,citecolor=blue,filecolor=blue,urlcolor=blue}
\usepackage[caption=false]{subfig}

\newcommand{\enzo}{\texttt{Enzo~}}
\newcommand{\music}{\textsc{Music~}}

\newcommand{\msolar} {$\rm{M_{\odot}}~$}
\newcommand{\msolarc} {$\rm{M_{\odot}}$}
\newcommand{\zsolar} {$\rm{Z_{\odot}}~$}

\newcommand{\molH} {$\rm{H_2}$~}
\newcommand{\molHc} {$\rm{H_2}$}

\newcommand{\ren} {\textit{Renaissance~}}
\newcommand{\pp} {\textit{Pop2Prime~}}
\newcommand{\ppc} {\textit{Pop2Prime}}
\def\Ol{$\Omega_\Lambda$}
\def\Om{$\Omega_{\rm m}$}
\def\Ob{$\Omega_{\rm b}$}

\begin{document}
\title[]{The Growth of Black Holes from Population III Remnants in the
  Renaissance Simulations}

\author[B.D. Smith, J.A. Regan, T.P. Downes, M.L. Norman,
  B.W. O'Shea \& J.H. Wise] 
{Britton D. Smith$^1\thanks{E-mail:bds006@sdsc.edu}$, John
  A. Regan$^2$\thanks{Marie Sk\l odowska-Curie Fellow},
  Turlough P. Downes$^2$, Michael L. Norman$^{1,3}$,\newauthor
  Brian W. O'Shea$^{4,5,6,7}$ \& John H. Wise$^{8}$\\
$^1$San Diego Supercomputer Center, University of California, San
  Diego, 10100 Hopkins Drive, La Jolla, CA 92093\\ 
$^2$Centre for Astrophysics \& Relativity, School of Mathematical
  Sciences, Dublin City University, Glasnevin, Ireland\\
$^3$Center for Astrophysics and Space Sciences, University of
  California, San Diego, 9500 Gilman Dr, La Jolla, CA 92093\\
$^4$National Superconducting Cyclotron Laboratory, Michigan State
  University, MI, 48823, USA\\
$^5$Department of Physics and Astronomy, Michigan State University,
  MI, 48823, USA\\
$^6$Department of Computational Mathematics, Science and Engineering,
  Michigan State University, MI, 48823, USA\\
$^7$Joint Institute for Nuclear Astrophysics - Center for the
  Evolution of the Elements, USA\\
$^8$Center for Relativistic Astrophysics, Georgia Institute of
  Technology, 837 State Street, Atlanta, GA 30332, USA
}

\pubyear{2016}
\label{firstpage}
\pagerange{\pageref{firstpage}--\pageref{lastpage}}

\maketitle

\begin{abstract} 
  The formation of stellar mass black holes from the remnants of
  Population III stars provides a source of initial black hole seeds
  with the potential to grow into intermediate or, in rare cases,
  possibly super-massive black holes. We use the \ren simulation suite
  to follow the growth of over 15,000 black holes born into
  mini-haloes in the early Universe. We compute the evolution of the
  black holes by post-processing individual remnant Population III
  star particles in the \ren simulation snapshots. The black holes
  populate haloes
  from $10^6$ \msolar up to $10^9$ \msolarc. We find that all of the
  black holes display very inefficient growth. On average, the black
  holes increase their initial mass by a factor $10^{-5}$, with the
  most active black holes increasing their mass by approximately
  10\%. Only a single black hole experiences any period of
  super-Eddington accretion, but the duration is very short and not
  repeated. Furthermore, we find no correlation of black hole
  accretion with halo mass in the mass range sampled. Within most
  haloes, we identify clumps of cool, dense gas for which accretion
  rates would be high, but instances of black holes encountering these
  clumps are rare and short-lived. Star formation competes with black
  hole growth by consuming available gas and driving down accretion
  rates through feedback.  We conclude that the black holes born from
  Population III remnants do not form a significant population of
  intermediate mass black holes in the early Universe and will need to
  wait until later times to undergo significant accretion, if at all.
\end{abstract}

\begin{keywords}
Cosmology: theory -- large-scale structure -- first stars, methods: numerical 
\end{keywords}


\section{Introduction} \label{Sec:Introduction}
The existence of supermassive black holes (SMBHs) in the first billion
years of the Universe presents a significant challenge to our
understanding of the formation of the first compact objects in our
Universe. The earliest SMBHs observed have masses upwards of a billion
solar masses \citep[e.g.][]{Fan_2006, Mortlock_2011, Venemans_2013,
Wu_2015, Banados_2018}. The means by which black holes could grow to
be so massive so quickly represents a serious theoretical challenge.
Beginning with the spherical accretion model, the maximum accretion
rate for a black hole can be expressed as
\begin{equation}
  M(t) = M_{0}\ \rm{exp} \Big( \frac{1 - \epsilon_{r}}{\epsilon_{r}}
  \frac{t}{t_{Edd}} \Big ),
\end{equation}
where $t_{Edd} = 0.45 $ Gyr and $\epsilon_{r}$ is the radiative
efficiency. For a ``standard'' radiative efficiency of $\epsilon_{r} \sim
0.1$ and a black hole seed mass of $M_{0} = 10^2$ \msolarc, it takes
nearly 1 Gyr to grow to M(t) $\sim 10^9$ \msolarc. At present, there
exist two potential origin stories for the SMBHs that inhabit the
centres of massive galaxies and shine as bright quasars. The seeds of
massive black holes may have been ``light'' ($\sim10 - 1000$
\msolarc), beginning as the remnants of the first stars
\citep[e.g.][]{Madau_2001} or from the core collapse of dense stellar
clusters \citep{Gurkan_2004, Gurkan_2006, Devecchi_2008,
Katz_2015}.  Alternatively, the seeds may have been ``heavy'' ($\sim
10^{3} - 10^{5}$ \msolarc), being the end product of supermassive star
formation \citep[e.g.,][]{Hosokawa_2012, Hosokawa_2013, Woods_2017}.

The first (Population III) stars in the Universe form out of
metal-free gas with \molH as the primary coolant. Within a few hundred
million years after the Big Bang, Population (Pop) III stars will
begin forming in 10$^{6}$ \msolar haloes \citep{Tegmark_1997,
Yoshida_2003a}. The inefficiency of \molH
as the sole gas coolant results in initial stellar masses that are
tens to hundreds of times more massive than the sun \citep{Bromm_1999,
Abel_2000, Abel_2002, Bromm_2002, OShea_2007b, Turk_2009,
Clark_2011a, Hirano_2014}. This range of masses provides multiple
pathways for black hole formation, including core-collapse supernovae
or hypernovae \citep[11 \msolar $\le$ M $\le$ 40
\msolarc;][]{1995ApJS..101..181W, 2006NuPhA.777..424N} and
direct formation \citep[40 \msolar $\le$ M $\le$ 140 \msolar and M
$>$ 260 \msolarc;][]{2002ApJ...567..532H}, leading to a population of
light seeds that have been implanted into the building blocks of
galaxies. The expected large number density of Pop III stars in
the early Universe \citep{2009ApJ...694..879T, 2013ApJ...773..108C}
makes them natural candidates for the seeds of SMBHs.\\
\indent Accretion onto Pop III remnants has been investigated as a
pathway for forming SMBHs through both analytical
\citep[e.g.][]{Madau_2001} and semi-analytical mechanisms
\citep[e.g.][]{Tanaka_2008, Pezzulli_2016}. Even if only a small
fraction of the Pop III remnant black holes grow at the Eddington
rate, it would be enough to seed the entire population of SMBHs
observed in the Universe. Detailed numerical simulations have also
been used to study the initial conditions surrounding the black hole
which forms from a Pop III star. These simulations take into account
the radiation field generated by the Pop III star, the formation of an
HII region surrounding the star, and any associated supernova
explosion. The stellar radiation and supernova successfully evacuate
the gas from the host halo, resulting in black holes that are `` born
starving'' \citep{Whalen_2004, 2005ApJ...628L...5O, Johnson_2007,
Milosavljevic_2009}. Simulations following the evolution for up to 200
Myr after initial formation find that these black holes continue to
experience no significant growth \citep{Alvarez_2009}. However, these
simulations did not have sufficient dynamic range to follow the
subsequent mergers of remnant mini-haloes into larger atomic cooling
haloes in which the black holes may be able to experience significant
accretion events. \\
\indent \citet{2012ApJ...754...34J} simulate the growth of
100 \msolar black holes from Pop III remnants with and without
feedback from the accreting black hole. In the case of no feedback,
they find that growth is negligible for $\sim$80 Myr until the halo
reaches the atomic cooling limit, at which time growth increases
significantly. When feedback is included, \citet{2012ApJ...754...34J}
find that growth remains insignificant through the end of the
simulation at z = 10. However, \citet{Volonteri_2015} claim that the
early bottleneck in growth might also be alleviated by short periods
of super-Eddington growth that allow black holes to grow by several
orders of magnitude in only 10 Myr. If a black hole is able
to migrate into an environment where super-critical accretion becomes
possible, then growth to a SMBH mass within the timescale of
approximately 500 Myr becomes possible \citep{Lupi_2016,
Valiante_2016, Pezzulli_2017, Pacucci_2017}.\\
\indent Heavy seeds emerge from a rarer, more exotic channel where
unusually high accretion rates lead to the formation of a supermassive
star \citep{Begelman_2006, Begelman_2008, Schleicher_2013,
Hosokawa_2013, Woods_2017, Haemmerle_2017}. These direct collapse
black holes (DCBHs) are thought to form in pristine atomic cooling
haloes where \molH formation has been suppressed preventing the
formation of smaller Pop III stars \citep{Wise_2008a, Regan_2009b,
Regan_2009, Agarwal_2012, Agarwal_2013, Becerra_2015, Latif_2013a,
Latif_2013c, Regan_2014a, Agarwal_2015b, Regan_2016a,
Regan_2017}. DCBH scenarios have the distinct advantages of starting
from much larger masses than light seeds and also existing in
environments where significantly more fuel is likely to be present
\citep{Hosokawa_2015, Nakauchi_2017}. However, the very existence of
supermassive stars is debated and furthermore it is not clear whether
the DCBH scenario can provide a sufficient number density of black
holes to explain the existence of all SMBHs.\\
\indent In this work, we seek to quantify the range of possibilities
for the Pop III light seed scenario in the early Universe.  We follow
the growth of 15,000 black holes in the \ren simulations
\citep{Xu_2013, Xu_2014, OShea_2015, Xu_2016, Xu_2016b} over
approximately 300 Myr, three orders of magnitude in halo mass, and
three different large-scale galactic environments.  We supplement this
data set with 12 Pop III remnants from the \pp simulations
\citep{Smith_2015}, which have superior mass, spatial, and time
resolution. Both of these sets of simulations follow the formation
and evolution of individual Pop III stars using
radiation-hydrodynamics in a cosmological context. We examine the
growth rates of the total black hole population and identify the
commonalities of those that grow the most. We then focus on the halo
with the most black holes. Finally, we study how the black hole growth
rate is regulated by star formation. The layout of the paper is as
follows. In Section \ref{Sec:Sims}, we describe the simulations used
in this work. In Section \ref{Sec:BHFormation}, we discuss the methods
for modeling black hole growth. In Section \ref{Sec:Results}, we
present the results of the investigations described above. Finally, we
conclude with a discussion and summary in Section \ref{Sec:Discussion}.

\section{Simulation Suites} \label{Sec:Sims}
All simulations analyzed in this work were performed with the
open-source, adaptive mesh-refinement + N-body code, \enzo
\citep{Enzo_2014}.  \enzo has been used extensively to simulate
high-redshift structure, including the formation of Pop III stars 
\citep{2002Sci...295...93A, 2005ApJ...628L...5O, 2007ApJ...654...66O,
  2008ApJ...673...14O, 2009Sci...325..601T, 2010ApJ...725L.140T,
  2011ApJ...726...55T, 2012ApJ...745..154T}, low-metallicity stars
\citep{2007ApJ...661L...5S, 2009ApJ...691..441S, 2014ApJ...783...75M,
  Smith_2015}, and the first galaxies \citep{2012MNRAS.427..311W,
  2012ApJ...745...50W, 2013ApJ...773...83X, 2014MNRAS.442.2560W,
  2014ApJ...795..144C}.  The two suites of simulations used here are
described below.

\subsection{Renaissance Simulations} \label{Sec:Renaissance}
The Renaissance Simulations
have been well detailed previously in the literature \citep{Xu_2013,
Xu_2014, Chen_2014, Ahn_2015, OShea_2015, Xu_2016b, Xu_2016}, and
here we only summarize the simulation characteristics relevant to this
study.  All of the Renaissance simulations were carried out in a
comoving volume of (40 Mpc)$^3$, created with the \music
\citep{Hahn_2011} initial conditions generator.  The cosmological
parameters were set using the 7-year WMAP $\Lambda$CDM+SZ+LENS best
fit \citep{Komatsu_2011}: \Om~=~0.266, \Ol~=~0.734, \Ob~=~0.0449, $h =
0.71$, $\sigma_8 = 0.81$ and $n = 0.963$.  First, an exploratory
simulation with $512^3$ particles ($1.7 \times 10^7$ \msolar per dark
matter particle) was run to $z=6$.
Three regions of interest were then selected for re-simulation at
higher resolution, namely a rare-peak region, a normal region and a
void region.
To generate the three regions the initial, lower resolution, volume
was smoothed on a physical scale of 5 comoving Mpc, and regions of
high ($\langle \delta \rangle \equiv \langle \rho \rangle / (\Omega_M \rho_C)-1 \sim 0.68$)
- the rare peak; average ($\langle \delta \rangle \sim 0.09$) - the normal region; and
low ($\langle \delta \rangle \sim -0.26$) - the void region.
The comoving volumes of the three regions were 133.6, 220.5 and 220.5 Mpc$^3$,
respectively. Each simulated region was then re-initialized with a
further three nested grids for an effective resolution of 4096$^3$ and
a dark matter particle resolution of $2.9 \times 10^4$ \msolar{}
within the high-resolution region. During the simulation, further
adaptive refinement was allowed up to a maximum 12 levels,
leading to a maximum spatial resolution of 19 
comoving pc (1.2~proper parsecs at $z = 15$).  The simulations were
evolved to a final redshift $z=15, 11.6$ and 9.9 for the Rare-Peak,
Normal and Void realisation respectively.
The halo mass function is well-resolved down to $2 \times 10^6$
\msolar(70 particles per halo), and at the ending redshift, the
three realisations contained a total of 822, 758, 458 galaxies having at least
1,000 particles ($M_{\rm vir} \simeq 2.9 \times 10^7$\msolar) and $\sim 15,000$
Pop III remnant black holes (see Table \ref{tab:summary}).

The simulations include both self consistent Pop III and
metal-enriched star formation (Pop II) at 
the maximum refinement level and capture star formation
in haloes as small as $3 \times 10^6$ \msolar \citep{Xu_2013}.
Pop III star formation is selected if the metallicity is less
than $10^{-4}$ of the solar fraction in the highest density cell with
metal-enriched star formation proceeding otherwise. The functional
form of the IMF is a power-law with a slope of -1.3 with an
exponential cutoff above a characteristic mass of 40 \msolarc.  The
operational mass range of the IMF is 1 \msolar $\le$ M $\le$ 300
\msolar (see \cite{Wise_2012b} for 
additional details.)
Stellar feedback uses the {\sc Moray} radiative transport
framework \citep{WiseAbel_2011} for H ionizing photons. Lyman-Werner (LW)
radiation that dissociates \molH is modeled using an optically thin,
inverse square law profile, centered on all star particles.  At the
end of their main-sequence lifetimes, Pop III stars in the mass range,
11 \msolar $\le$ M $\le$ 40 \msolarc, explode as core-collapse
supernova with total energies and metal-yields
calculated by \citet{2006NuPhA.777..424N}. Pop III stars in the mass range
140 \msolar $\le$ M $\le$ 260 \msolar explode as pair-instability
supernova (PISN) with total energy of $\sim6-100\times10^{51}$ erg
over the PISN mass range and metal yields calculated by
\citet{2002ApJ...567..532H}. For Pop III stars outside of the above
mass ranges (40 \msolar $<$ M $<$ 140 \msolar and M $>$ 260 \msolarc),
no feedback is added after the main-sequence lifetime.
The ionization
states of hydrogen and helium are followed with a 9-species primordial
non-equilibrium chemistry and cooling network \citep{Abel_1997},
supplemented by metal-dependent cooling tables \citep{Smith_2009}. No
\molH self-shielding in included in the simulations as the densities
at which it becomes relevant are not fully resolved by the simulations. 
A LW background radiation field is also included to model radiation
from stars which are not within the simulation volume \citep{2012MNRAS.427..311W}.
In the high density region of the rare-peak
simulation, the LW radiation from stars dominates over the background.
Although the simulations cannot follow Pop III star formation in haloes below
$3 \times 10^6$ \msolarc, star formation is suppressed by the LW background in such
haloes \citep{Machacek_2001, Wise_2007b, OShea_2008}.
Finally, we note the existence of multiple versions of each of these
three simulations in the literature. The variations were run to
different redshifts and used slightly different methods for calculating
the global Lyman-Werner radiation field. For clarity, we list
the simulations used here in Table \ref{tab:summary} by the names
given to them in the upcoming public data release along with the
reference of their first appearance.

\begin{table}
\caption{Simulation Summary \label{tab:summary}}
\begin{tabular}{lcccc}
\hline
Simulation$^a$ & V$^{b}_{\rm hr}$ [(Mpc com.)$^{3}$] & z$_{\rm
  f}^{c}$ & N$_{\rm black\ holes}^{d}$ & Ref.$^{e}$\\
\hline
Rare-Peak\_LWB & 133.6 & 15.0  & 6518 & 1 \\
Normal\_BG1    & 220.5 & 11.6  & 6225 & 2 \\
Void\_BG1      & 220.5  & 9.9  & 2487 & 3 \\
Pop2Prime      & 0.004 & 10$^{*}$ & 12  & 4 \\
\hline
\end{tabular}
\\ (a) the simulation name; (b) volume of the high resolution region;
(c) the final redshift of the simulation; (d) the total number of Pop
III remnant black holes at the final redshift. * - after $z = 11.83$,
this simulation was continued to $z = 10$ with star formation turned
off; (e) publication of first appearance.  1: \citet{OShea_2015}, 2:
\citet{Xu_2016}, 3: \citet{Xu_2016b}, 4: \citet{Smith_2015}.
\end{table}

\subsection{Pop2Prime Simulations} \label{Sec:PopIIPrime}

We supplement the \ren simulations with an extension of the simulation
presented in \citet{Smith_2015}, referred to here as the \pp
simulation.  With significantly higher mass and spatial resolution,
the \pp simulation provides some constraint on the dependence of the
results on resolution.  The \pp simulation uses a 500 comoving kpc/h
box, initialized at $z = 180$ with the \music initial conditions
generator with the WMAP 7 best-fit cosmological parameters,
$\Omega_{m} = 0.266$, $\Omega_{\lambda} = 0.732$, $\Omega_{b} =
0.0449$, H$_{0} = 71.0$~km/s/Mpc, $\sigma_{8} = 0.801$, and n$_{s} =
0.963$ \citep{2011ApJS..192...18K}, and using a
\citet{1999ApJ...511....5E} transfer function and second-order
Lagrangian perturbation theory.  The simulation follows the region
around a halo reaching a virial mass of 1.7 $\times10^{7}$ \msolar at
$z = 10$.  The initial conditions are generated with 512$^{3}$ grid
cells and dark matter particles on the root grid and two additional
levels of nested refinement surrounding the target halo, corresponding
to a comoving spatial resolution of 0.244~kpc/h, and a baryon (dark
matter) mass resolution of 0.259~\msolar (1.274~\msolarc).

The \pp simulation includes the formation and feedback from Pop III
stars in a manner similar to the \ren simulations, with the addition
of He ionizing radiation (only H ionizing radiation was used in the
\ren simulations) using the {\sc Moray} adaptive ray-tracing
method and treating LW radiation as optically thin with 1/$r^2$
attenuation.  In contrast to the \ren simulations, which adopt a
power-law Pop III IMF, all Pop III stars in \pp are given a mass of 40
\msolar and end their main-sequence lifetimes (3.86 Myr) in a
core-collapse supernova with total energy of 10$^{51}$ erg. Since the
original goal of the \pp simulation was to study the collapse and
fragmentation of metal-enriched gas, this simulation does not form any
Pop II stars.  Instead, gas with metallicity greater than 10$^{-4}$
\zsolar is allowed to collapse until a number density of $\sim10^{13}$
cm$^{-3}$ is reached, at which time the simulation stops.  This occurs
at $z \sim 11.8$ after a total of 12 Pop III stars have formed.  The
simulation is then carried forward to $z = 10$, roughly an additional
100 Myr, with star formation turned off.  This serves as an
illuminating experiment of the effects of stellar feedback on black
hole growth.  The \pp simulation uses the same chemistry and cooling
machinery as the \ren simulations, but with the additions of three
deuterium species (D, D$^{+}$, and HD), \molH formation on dust
grains \citep[described in][]{2014ApJ...783...75M}, and self-shielding
of LW radiation using the model of \citet{Wolcott-Green_2011}.

\section{Analysis}

\subsection{Black Hole Formation in the
  Simulations} \label{Sec:BHFormation}
The simulations discussed here do not contain a subgrid prescription
for black hole formation.  Regardless of their initial mass, Pop III
star particles are given a negligible mass at the end of their
main-sequence lifetimes and do not accrete from their
surroundings. They then effectively act as extremely low mass dark
matter particles.  The simulations continue to update the positions,
velocities, and accelerations of these particles due to gravity, so
they will be located in approximately the same positions as if they had
been evolved fully self-consistently as black holes.  We rely on
this fact to model the growth of black holes represented by these
particle using accretion rates calculated from the local gas
conditions within the available simulation snapshots.  This is not
self-consistent and completely ignores the gravitational force of the
black holes on the surrounding material, changes in
momentum/trajectory of the particle due to accretion, and radiative
feedback from accretion.  Nevertheless, this exercise can provide a rough
estimate of the evolution of stellar mass black holes in the early
universe across different galactic environments.  Within the
uncertainties created by the first two caveats, the lack of radiative
feedback serves to provide an upper limit on the overall black hole
growth.

\subsection{Calculating Black Hole Growth} \label{Sec:BHAnalysis}

We compute growth histories for each Pop III star particle expected to
form a black hole based on its zero-age main-sequence (ZAMS) mass.
Given the initial mass of the star particle, we calculate its initial
black hole mass by linearly interpolating from the results of
\citet[][Table 3]{1995ApJS..101..181W} for stars with M $<$ 140
\msolarc.  For stars with 140 \msolar $<$ M $\le$ 260 \msolarc, the star
has undergone a PISN and so we assume no compact remnant.  For
stars with 260 \msolar $<$ M $\le$ 300 \msolar (the upper limit of the
Pop III IMF in the \ren simulations), we set the initial black hole
mass to be the mass of the He core using the relation from
\citet[][Equation 1]{2002ApJ...567..532H}, given by
\begin{equation}
M_{\rm He} \simeq \frac{13}{24}\ (M_{*} - 20\ \rm{M}_{\odot}).
\end{equation}

We model black hole growth as spherical Bondi-Hoyle accretion
\citep{1941MNRAS.101..227H, 1952MNRAS.112..195B}, where the growth rate is given by
\begin{equation} \label{Bondi}
  \dot{m}_{B-H} \simeq \frac{\alpha \pi \rho G^2 M^{2}_{BH}}{{\rm
      max}(|\vec{v}|, c_s)^3},
\end{equation}
where $M$ is the mass of the black hole, $\rho$ is the gas density
surrounding the black hole, $c_{s}$ is the local sound speed,
$|\vec{v}|$ is the magnitude of the velocity of the black hole
relative to the surrounding material, and $\alpha$ is a dimensionless
boost factor.  \citet{Krumholz_2005} show that the accretion rate is
decreased when the gas has non-zero vorticity, but we ignore this
effect to consider the most optimistic growth scenario.  The boost
factor term was first added to Equation \ref{Bondi} by
\citet{2005MNRAS.361..776S} to account for underestimation in the gas
density in the vicinity of the black hole caused by limited spatial
resolution of the simulation.  This scale is the Bondi radius, given
by
\begin{equation}
r_b = \frac{2 G M_{BH}}{c_{s}^2},
\end{equation}
which we do not resolve in our
simulations. \citet{2009MNRAS.398...53B} excellently 
summarize the subsequent use of the boost factor in proceeding works,
noting the commonly adopted constant values of $\alpha = 100-300$
\citep[although see][for alternative approaches to the boost
  factor]{2007ApJ...665..107P, 2011ApJ...738...54K}.
However, \citet{2009MNRAS.398...53B} argue that values of $\alpha > 1$
are unphysical when the medium is single-phase and its associated
Jeans length is resolved by the simulation.  In the \pp simulations,
the Jeans length is resolved explicitly by a minimum of 64 grid
cells.  In the \ren simulations, Jeans length-based refinement is not
used, but we find that in practice the Jeans length is refined in the
vicinity of the black hole particle by at least 4 cells (i.e., the grid cell containing
it) roughly 99.9\% of the time.  \citet{2009MNRAS.398...53B} argue
that boost factors should be used when the medium is expected to be
multi-phase and the associated spatial scales are unresolved.  The
scale of the multi-phase medium is not set by the Jeans length, but
instead by the cooling length (the cooling time multiplied by the
sound speed) as it forms through thermal instability
\citep{2017ApJ...845...80V, 2018MNRAS.473.5407M}.  For the
circumgalactic medium, \citet{2017MNRAS.466.3810F} find that gas is
stable against going multi-phase for halo masses below roughly
10$^{11.5}$ \msolarc, relating the associated virial temperature to the
point in the cooling curve where cooling times become long.  In our
case, we find that the black hole particles spend the majority of
their time in two regimes: hot ($T \sim 10^{6-7}$ K), underdense ($n <
10^{-2}$ cm$^{-3}$) gas that is the product of stellar feedback; and
cooler ($T \sim 10^{4}$ K), denser (10$^{-2}$ cm$^{-3}$ $<$ n $<$ 10$^{2}$
cm$^{-3}$) gas heated to the virial temperature, but unable to cool
further due to its low
metallicity. Due to the long cooling times, we expect that in practice
the thermal instability will have no impact and hence the gas will be
single phase in both these regimes.  Therefore, we choose to adopt a constant value of
$\alpha = 1$ (i.e., no boost) in our growth model.  Finally, we do not
cap the black hole growth rate at the Eddington limit, which is given
by
\begin{equation}
\dot{m}_{Edd} = \frac{4 \pi G M_{BH} m_{p}}{\epsilon_{r} \sigma_{T} c}
\simeq 2.2 \times 10^{-8} \left( \frac{0.1}{\epsilon_{r}} \right)
\left( \frac{M_{BH}}{M_{\odot}} \right) [M_{\odot} / yr],
\end{equation}
where $m_{p}$ is the proton mass, $\epsilon_{r}$ is the radiative
efficiency, $\sigma_{T}$ is the Thomson cross-section, and $c$ is the
speed of light.  Throughout this work, we refer to the Eddington rate
assuming $\epsilon_{r} = 0.1$, appropriate for a non-rotating
Schwarzschild black hole \citep{Shakura_1973}. As we show below,
instances of near-Eddington accretion are extremely rare. We allow for
super-Eddington accretion only to highlight instances where the
physical conditions create a situation where it could be possible.

Starting with the first simulation snapshot after which a Pop III star
particle has exceeded its main-sequence lifetime, we use Equation
\ref{Bondi} to compute the particle's instantaneous growth
rate\footnote{All analysis codes used in this work, including
  figure-generating scripts, are available as an extension package for
  the \texttt{yt} analysis code \citep{YT}
  at \url{https://github.com/brittonsmith/yt_p3bh}.}.
Assuming the density of the grid cell decreases negligibly due to
accretion by the particle, Equation \ref{Bondi} can be solved
analytically to give the black hole's mass at snapshot $i+1$, given
its mass at snapshot $i$, $M_{i}$, and the timestep between snapshots,
$\Delta t$ as
\begin{equation}
M_{i+1} = \frac{M_{i}}{1 - \frac{\dot{m}_{B-H} \Delta t}{M_{i}}}.
\end{equation}
We compute the final mass of each black hole particle by iterating
over all available snapshots for each simulation.  For the \ren
simulations, the average time between snapshots is roughly 4 Myr.  For
the \pp simulation, the average time between snapshots is about 0.8 Myr.

\begin{figure}
    \includegraphics[width=0.5\textwidth]{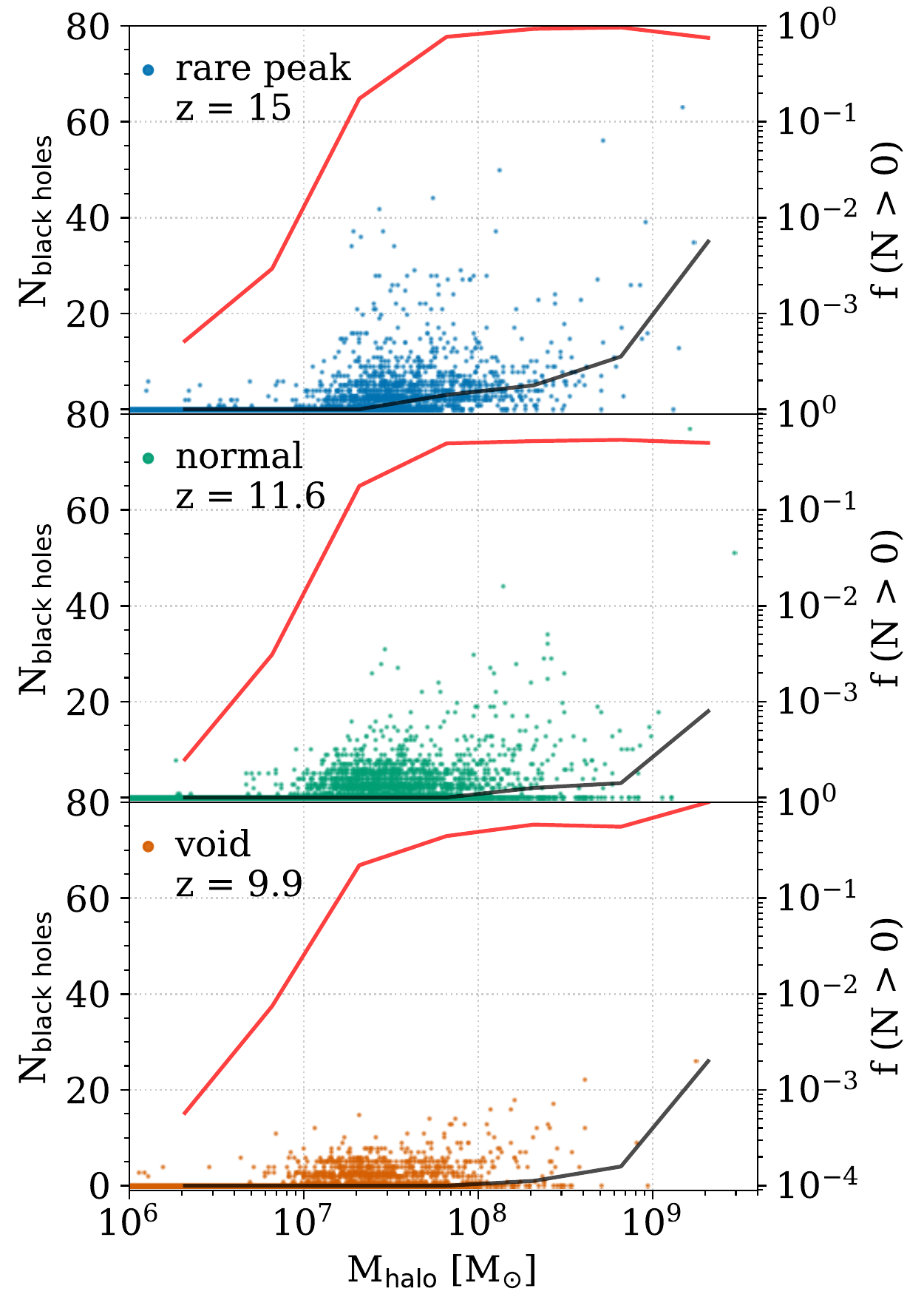}
    \caption[] {\label{BHHalo} 
      Number of black holes in a halo as a function of halo mass for
      the rare peak (top), normal (middle), and void (bottom)
      runs of the \ren simulations.  The black line indicates the
      median in bins of 0.5 dex and the red line shows the fraction of
      haloes with at least one black hole.}
\end{figure}

\begin{figure*}
    \includegraphics[width=\textwidth]{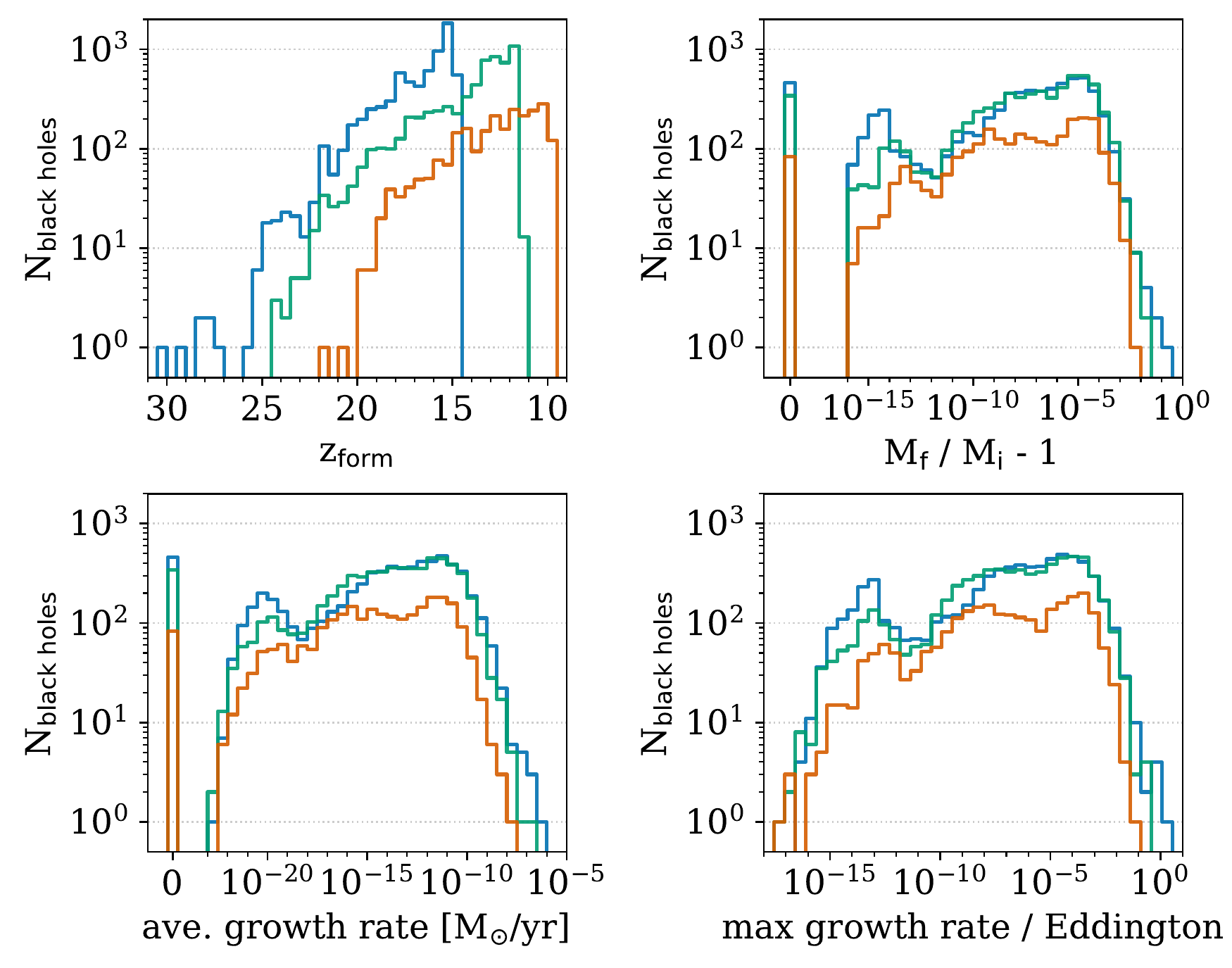}
    \caption[]
     { \label{BlackHoleGrowthPanel}
       Black hole population statistics for the final snapshot of the
       rare peak (blue, $z = 15$), normal (green, $z = 11.6$), and
       void (orange, $z = 9.9$) runs of the \ren simulations.
       \textit{Top Left Panel}: black hole formation redshift. The
       drop-off corresponds to the final redshift of each simulation.
       \textit{Top Right Panel}: relative overall black hole growth. 
       The vast majority of black holes grow by a neglible amount. No
       black holes are able to increase their mass by more than 10\%.
       \textit{Bottom Left Panel}: average absolute growth rate.
       \textit{Bottom Right Panel}: maximum instantaneous growth rate,
       as a fraction of the Eddington rate, achieved at any point
       during the simulation.}
\end{figure*}

\section{Results} \label{Sec:Results}
Below, we present the results of growing the Pop III
remnant black holes for each simulation to its final snapshot,
focusing primarily on the \ren simulations. We note that the Rarepeak
and Normal simulations were run until a qualitatively similar amount
of structure (number of haloes, stars, etc.) had formed, hence the
similarity in the number of black holes formed. However, this
is not true for the Void simulation, whose final redshift was
determined by a prior simulation that was used to create a LW
background model for the Void simulation. Unless otherwise stated, the
results shown refer to the final output of each simulation. Table
\ref{tab:summary} lists the final redshift and total number of black
holes formed in each simulation.

\begin{figure}
    \includegraphics[width=0.5\textwidth]{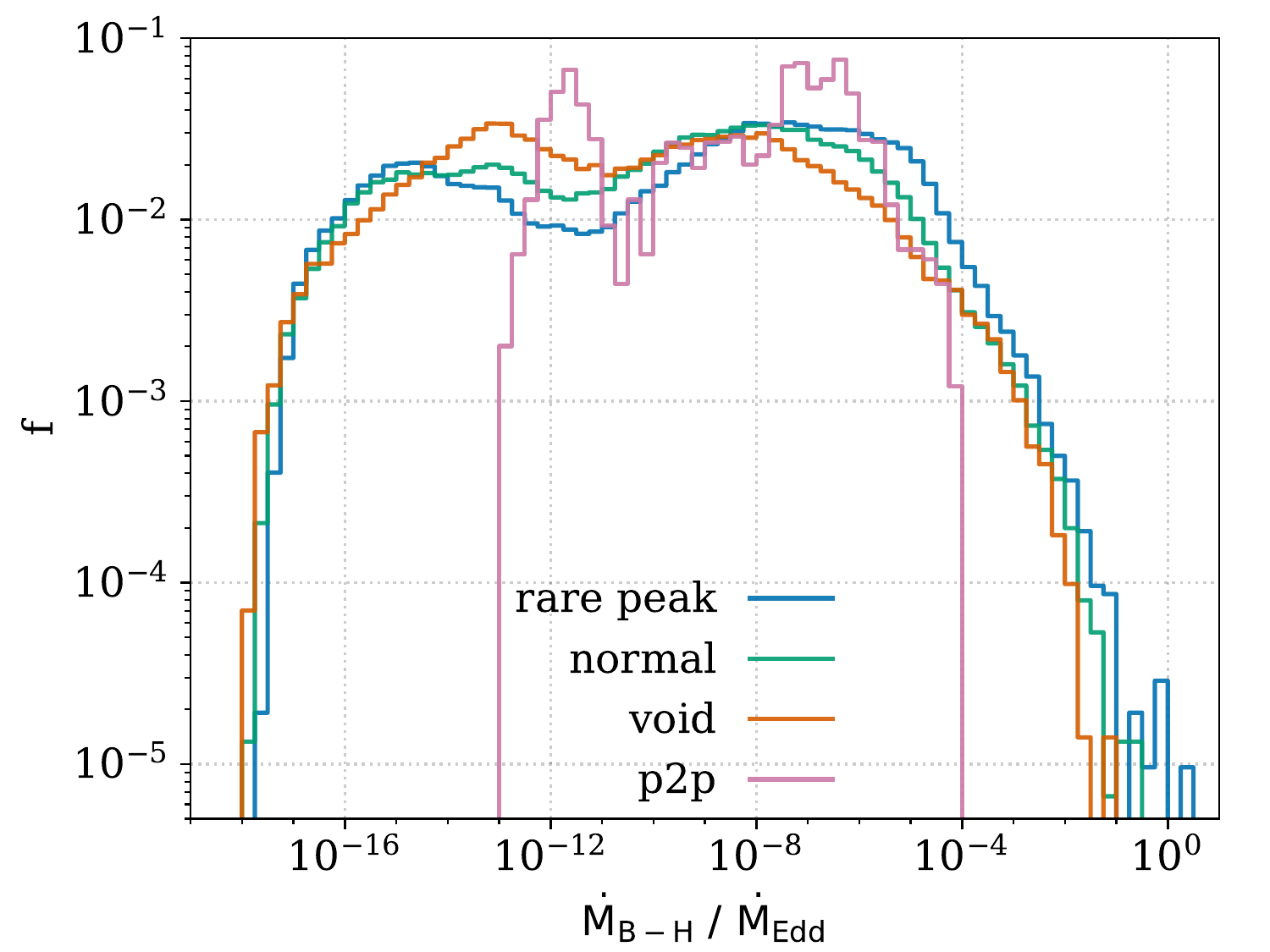}
    \caption[]
            {\label{all_growth}
              Probability distribution function of all instantaneous
              growth rates for all black holes over all times for all
              simulations. Growth rates are shown as a fraction of the
              Eddington rate. For the \pp simulation (p2p above), the period
              of time where star formation is turned off is not shown
              as this is unphysical.
             }
\end{figure}

\begin{figure}
    \includegraphics[width=0.5\textwidth]{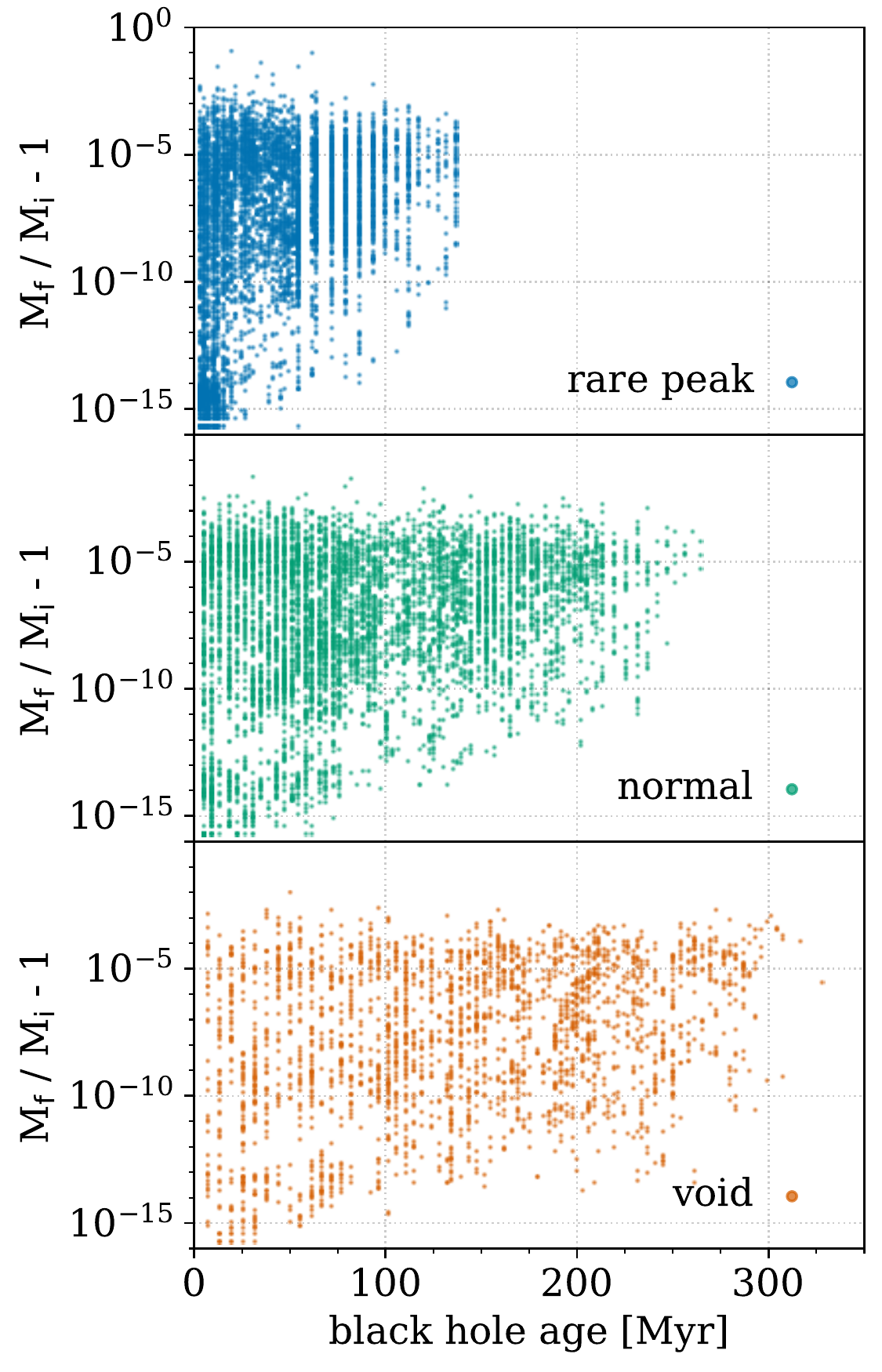}
    \caption[]
            {\label{BHGrowthCorrelation} Relative black hole growth as
              a function of black hole age, where M$_{i}$ and M$_{f}$
              are the initial and final masses, respectively. No
              strong trend of black hole growth versus age exists.
              However, black holes with the most and least growth are
              generally young ($\lesssim 50$ Myr).
             }
\end{figure}
\subsection{Where are the black holes?}
In Figure \ref{BHHalo}, we show the number of black holes as a
function of halo mass at the final snapshot of each of the \ren
simulations. Over the three simulations we find that the distribution
of black holes is scattered from haloes as small as a few times $10^6$
\msolarc, roughly the resolution limit, up to approximately $10^9$
\msolarc, the maximum halo mass. The void haloes (bottom panel) show
the smallest number of black holes per halo with on average less than
one black hole per halo up to M$_{halo} \sim 10^7$ \msolarc. The
normal and rare haloes show a slightly larger scatter with a handful
of haloes having up to 40 black holes per halo up to M$_{halo} \sim
10^7$ \msolarc. In all cases, less than roughly 1\% of haloes with
M$_{halo}$ $<$ 10$^7$ \msolar are populated with Pop III black holes.
Above a mass of M$_{halo} \sim 10^7$ \msolarc, the 
number of black holes shows a marked increase in number, especially
for the normal and rare peak haloes. This is because at this mass
scale atomic haloes form through the merger and accretion of 
mini-haloes which previously hosted black holes. \cite{Xu_2013} found
in their investigation of Pop III stars in the Renaissance simulations
that the number of Pop III stars and remnants peaks in 
haloes with masses of a few $\times 10^7$ \msolar and that the growth
in the number of remnants comes mainly from mergers of
mini-haloes. They find that Pop III stars form only in haloes with
masses between $4 \times 10^6$ and $3 \times 10^8$ \msolarc. Pop III
stars found in higher mass haloes appear there via mergers. It is the
remnants of these Pop III stars that now populate the more massive
atomic cooling haloes. The median line (solid black line in Figure
\ref{BHHalo}) shows a clear increase in the number of black holes in
haloes more massive than approximately $10^8$ \msolar due to the
effect of mergers. We therefore sample quite well the black hole
occupation fraction in haloes up to M$_{halo} \sim 10^9$ \msolarc.\\

\subsection{How much do the black holes grow?} \label{growth}
In Figure \ref{BlackHoleGrowthPanel}, we show histograms of black hole
properties, including formation redshift, relative growth, final mass,
and maximum instantaneous growth rate. As expected, the black hole
formation rate is indicative of the large-scale overdensity associated
with each simulation.
The landscape of overall black hole growth
is notably bleak. Not a single black hole is able to double in
size, with the best cases growing by roughly 13\% in the Rarepeak.
In the less dense galactic environments, the maximal mass growth is
even lower, with the best case in the Normal run growing only by 2\%,
and that of the Void run by just under 1\%. In all three cases, the
distribution of relative black hole
growth is bimodal, with peaks at $\sim10^{-15}$ to $\sim10^{-13}$ and
a broader peak from $\sim10^{-8}$ to $\sim10^{-3}$. The distribution
of maximum instantaneous growth rates closely resembles the overall
relative growth.  In all, only a single black hole in the Rarepeak is
able to achieve super-Eddington accretion. The overwhelming majority
of black holes accrete maximally at less than 10$^{-4}$ of their
Eddington rates. Figure \ref{all_growth} shows a probability
distribution function of all instantaneous growth rates for all black
holes and all snapshots.  Only 2-3\% of all growth rates exceed
10$^{-4}$ of the Eddington rate the total number of super-Eddington
events is just one, i.e., the one black hole that experiences
super-Eddington growth does so only once.

The \pp simulation shows a similar
bimodal distribution of individual growth rates, albeit with narrower
peaks and an overall much smaller range of total values. The two peaks
correspond to two dinstinct physical conditions in which the black
holes tend to exist. The lower of the two peaks is from hot,
underdense gas associated with stellar feedback. All black holes
forming in a supernova event will live in this phase at least once,
and likely much longer given the long associated cooling times and
continually occurring star formation. The higher peak comes from gas
about to form stars, where the medium is slighly denser and heated to
roughly the virial temperature. In \ppc, the lower peak occurs at a
higher growth rate because of the relative weakness of the stellar
feedback producing lower temperatures in the hot phase. Haloes in \ppc,
with masses of only a few hundred thousand \msolarc, form only 1-2 stars
total. These smaller haloes also have lower virial temperatures and
central gas densities, thus moving the location of the second peak in
Figure \ref{all_growth} to lower accretion rates.

\begin{figure}
    \includegraphics[width=0.5\textwidth]{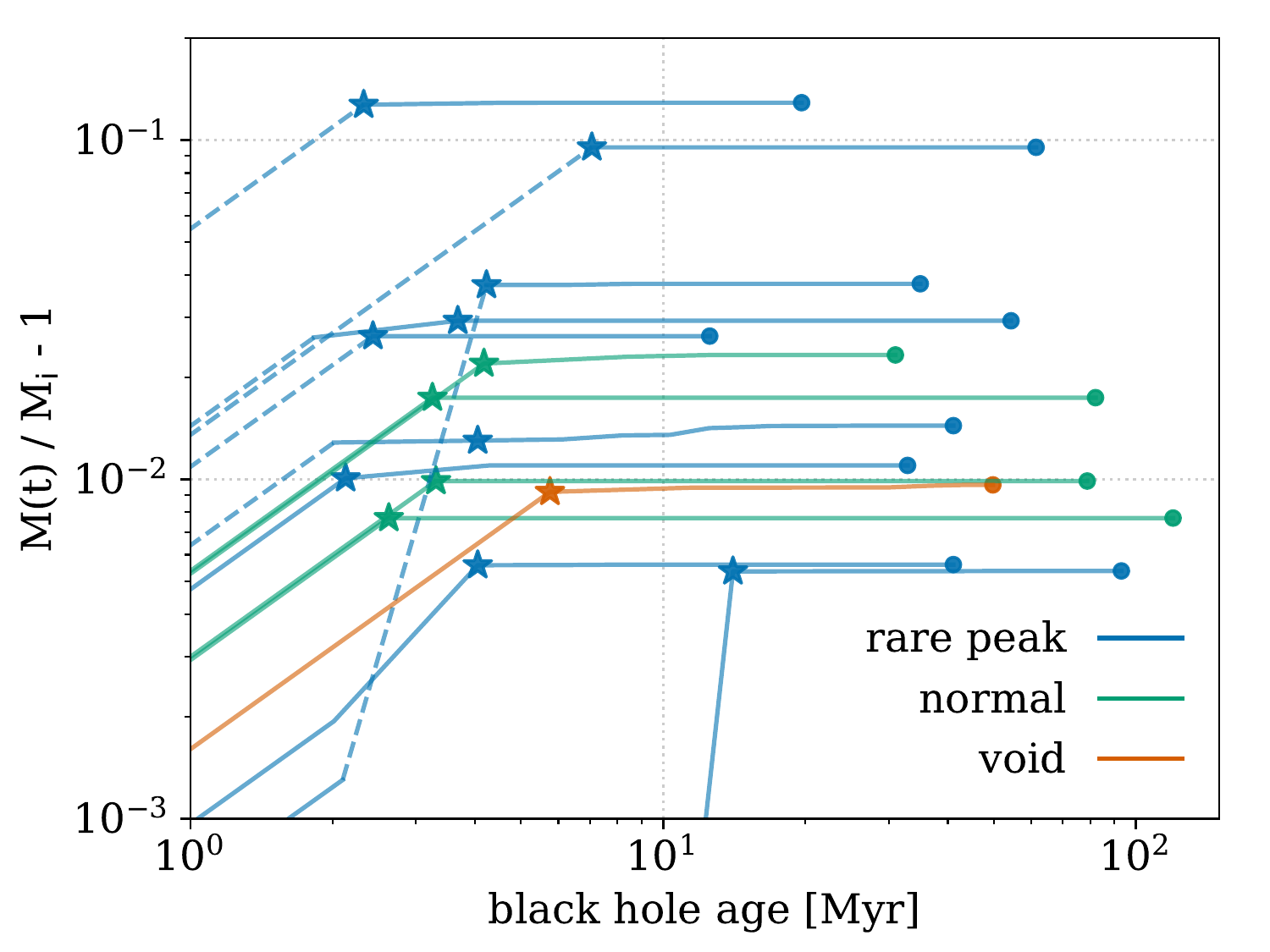}
    \caption[]
            {\label{growthhistory}
              Relative black hole mass as a function of age for all
              black holes with total relative growth of at least
              0.005.  Stars indicate the time when the black hole has
              accomplished 90\% of its total growth and the circles
              denote its final mass and age.  Dashed lines indicate
              periods of black hole growth of at least 0.25 of the
              Eddington rate.
             }
\end{figure}

In Figure \ref{BHGrowthCorrelation}, we plot the relative growth of
black holes as a function of their age. For black holes in the bulk of
the relative growth distribution (Figure \ref{BlackHoleGrowthPanel},
top-right panel), there is effectively no relation between overall
growth and age apart from the lack of older black holes at the lowest
values of relative growth. The black holes showing the
most growth are relatively young, with ages less than about 50 Myr. In
Figure \ref{growthhistory}, we show the individual growth histories
for all black holes growing by at least 0.5\%. In all but one case, these
black holes reach $>90\%$ of their final mass in less than 10 Myr. 
There are 9 black holes in the Rarepeak realisation that grow by at
least 0.5\%, 4 in the Normal, and only one in the Void. Of the 14
black holes shown here, 6 reach accretion rates of at least one
quarter of Eddington, with one reaching 2.5 times Eddington.  In all
cases, this strong growth lasts for only a single snapshot, and is
therefore likely overestimated. Not surprisingly,
all 14 of these black holes are in the mass windows where formation
occurs without a preceding supernova, i.e., 40 \msolar $<$ M $<$ 140
\msolar and M $>$ 260 \msolarc. These stellar mass ranges correspond
to initial black hole masses of 16.6 \msolar $<$ M $<$ 65 \msolar and
M $>$ 130 \msolarc. Apart from this initial period of
super critical growth, no black holes are able to accrete at rates
exceeding the Eddington limit, with most accreting at rates many
orders of magnitude below the Eddington rate.

\begin{figure}
    \includegraphics[width=0.5\textwidth]{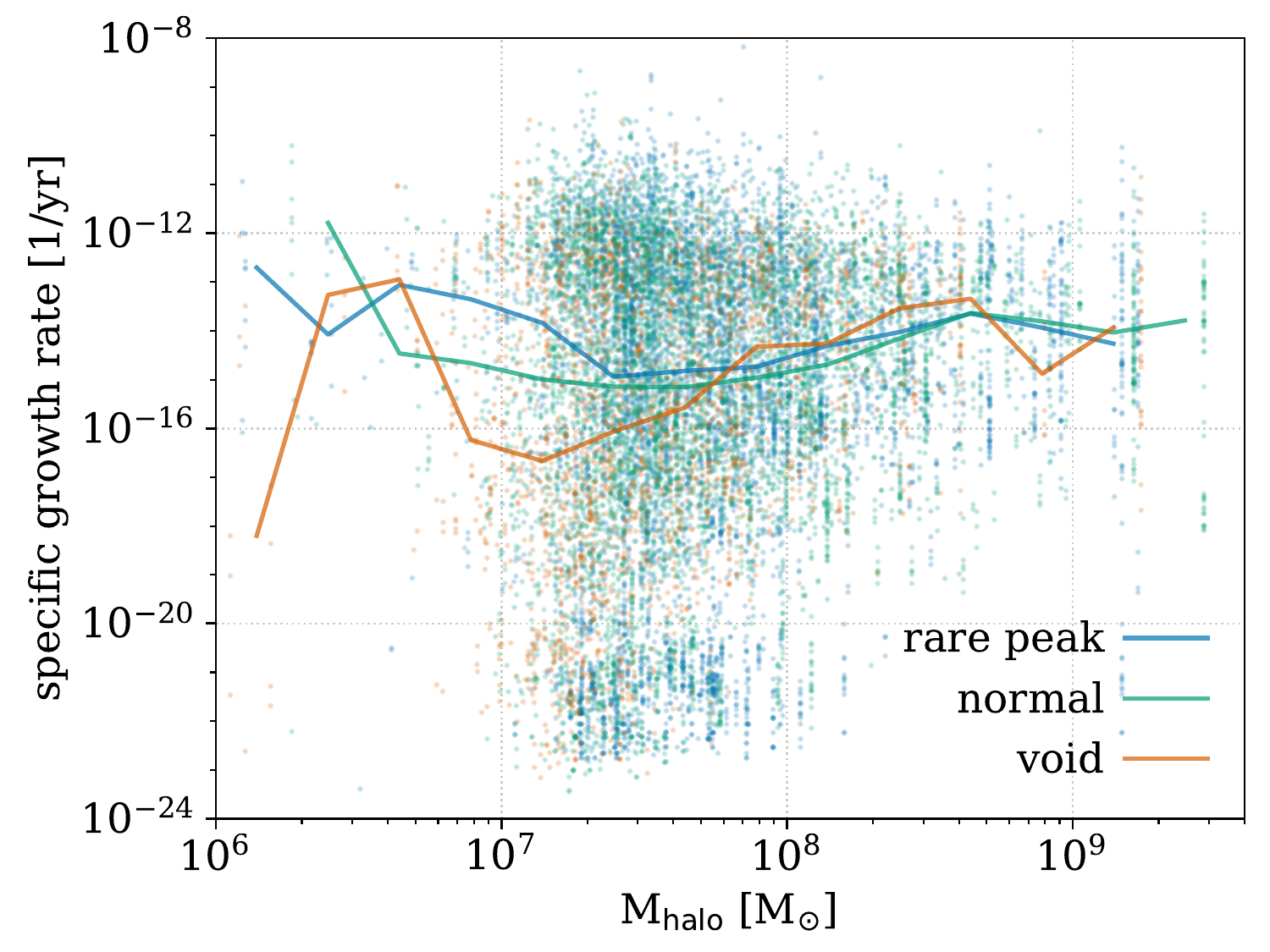}
    \caption[]
            {\label{BlackHoleGrowth}
              Specific black hole growth rate as a function of halo
              mass, where specific growth rate is defined as (M$_{f}$
              - M$_{i}$) / (M$_{i}$ $\times$ age). Solid lines indicate
              median number of black holes per halo in mass bins of
              0.25 dex.  We find no 
              correlation of black hole growth with halo mass up to
              $\rm{M_{halo} \sim 10^9}$ \msolarc.
             }
\end{figure}

Finally, in Figure \ref{BlackHoleGrowth} we plot the specific growth
rate (average growth rate divided by initial mass) for all black holes
as a function of host halo mass. Within the mass range tracked by the
\ren simulations (10$^{6}$ \msolar $<$ M$_{halo}$ $<$ 10$^{9}$
\msolarc), we see no evidence of the larger gas reservoirs of more
massive haloes aiding in black hole growth, except to the extent that
higher mass haloes show a scarcity of the most
slowly growing black holes.  This finding appears to be in agreement
with the isolated galaxy simulations of \citet{2007ApJ...665..107P},
who find no instances super-Eddington growth in haloes up to 10$^{10}$
\msolarc. The cosmological simulations of \citet{2017MNRAS.468.3935H},
which include black hole growth with feedback in a 10 Mpc comoving
box, also find very limited accretion at high redshift. Similar to
this work, the distribution of accretion rates in
\citet{2017MNRAS.468.3935H} also show a peak around 10$^{-5}$ of the
Eddington rate, although their larger box size and lower final
redshift (at the cost of lower resolution) are able to capture more
instances of much higher accretion rates.

The global statistics shown in Figures \ref{BHGrowthCorrelation},
\ref{growthhistory} and \ref{BlackHoleGrowth} support the conclusion
that light seeds, born from Pop III remnant black holes do not grow
efficiently by accretion in haloes up to $\rm{M_{halo} \sim 10^9}$
\msolarc. Next, we examine a single halo in detail to understand why
these black holes are unable to grow.

\begin{figure}
    \includegraphics[width=0.5\textwidth]{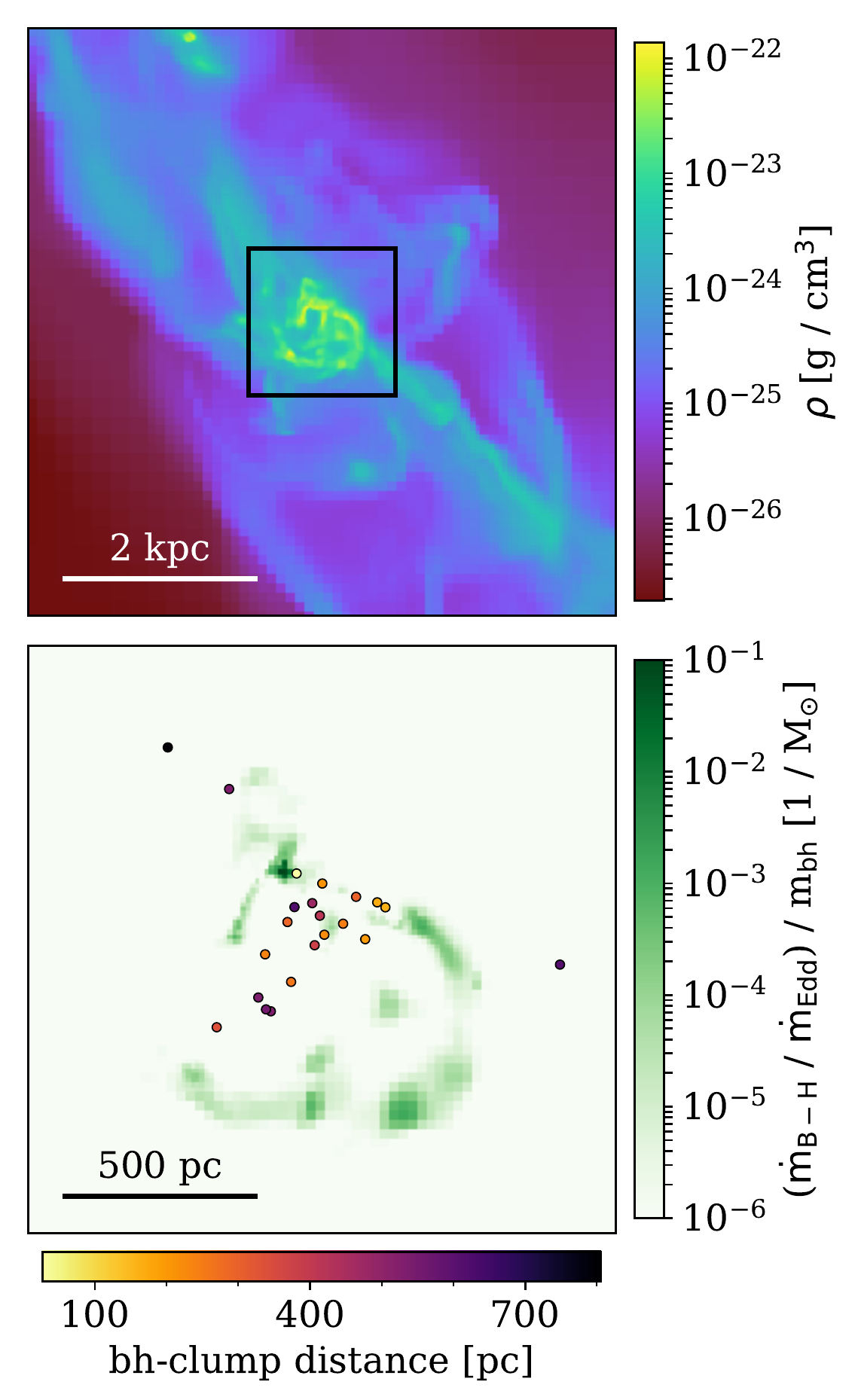}
    \caption[Halo from normal region with most black holes.]
            { \label{projection} \textit{Top Panel:} projection of
              mass-weighted mean density for the halo with the most
              black holes (M$_{\rm halo}$ = 1.6$\times10^{9}$ \msolarc,
              $n_{BH} = 77$).  The size of the projected region
              denotes the halo's virial radius of 3.4 kpc.
              \textit{Bottom Panel:} mass-weighted projection of gas
              accretability, defined as the ratio of Bondi-Hoyle to
              Eddington, divided by black hole mass. For example, a
              value of 0.1 M$^{-1}_{\odot}$ would
              allow a 1 \msolar black hole to accrete at 0.1 Eddington
              and a 10 \msolar black hole to accrete at Eddington.
              The projected region is 0.25 of the virial radius.
              Circles indicate the locations of black holes, with
              colors denoting the distance to the nearest gas
              clump with an accretability of at least
              10$^{-3}$ M$^{-1}_{\odot}$.
            }
\end{figure}

\begin{figure}
    \includegraphics[width=0.5\textwidth]{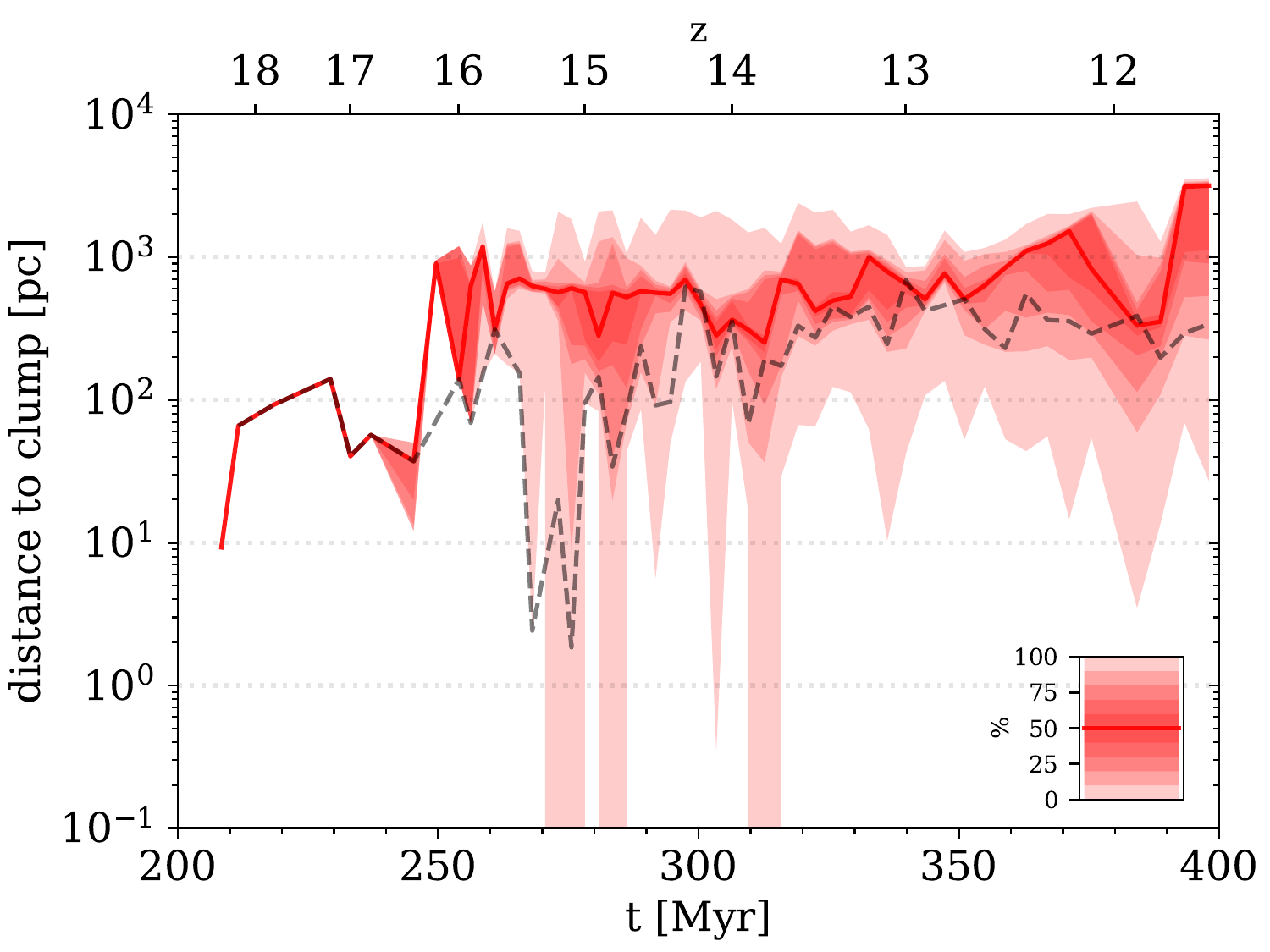}
    \caption[]
            { \label{fig:distances} Distribution of distances between
              a black hole and a high accretability gas clump
              (greater than 10$^{-3}$ M$^{-1}_{\odot}$)
              for all progenitor haloes of the halo hosting the most
              black holes (shown in Figure \ref{projection}.)  Where
              the shaded region extends to the bottom of the figure,
              some black holes exist within high accretion rate clumps.
              See Table \ref{tab:bhclump} for a list of all instances
              of black holes within clumps.  Values shown in red
              correspond to all black holes within a halo's virial
              radius.  The black, dashed line denotes the median
              separation for black holes within 0.25 of the virial radius.
            }
\end{figure}

\subsection{Why do the black holes not grow?} \label{whynot}
As we have seen the black hole accretion rate shows no marked increase
as a function of time or halo mass. To further understand the
evolution of Pop III remnant black holes we examine in detail the halo
with the most black holes, coming from the Normal run with a total of
77 black holes at the final output. In Figure \ref{projection}, we plot the halo's
large-scale gas distribution with the effective ``accretability'' of
the central, dense gas shown in the bottom panel. Here, we define the
accretability as the ratio of the Bondi-Hoyle rate to the Eddington
rate, divided by black hole mass. If we ignore the relative motion
term in Equation \ref{Bondi} and consider only the sound speed of the
gas, the above quantity is independent of black hole mass and is
simply a measure of the gas properties. Accretability has units of
M$^{-1}_{\odot}$, meaning that for an accretability of 0.1 M$^{-1}_{\odot}$, a 1
\msolar black hole would accrete at 0.1 of Eddington and a 10 \msolar
black hole would accrete at Eddington. Regions of high accretability
are clearly associated with dense gas, but the converse of that
statement is not necessarily true. Interestingly, high accretability
clumps do not appear to be centrally concentrated.

In the bottom panel of Figure
\ref{projection}, we overplot the locations of all black holes in the
inner halo, ($r < 0.25\ r_{vir}$.) We use the \texttt{yt} clump finder
\citep{Smith_2009, YT} to
identify all topologically disconnected regions with accretability of
at least $10^{-3}$ M$^{-1}_{\odot}$. We choose this value as it is the minimum
value for a few hundred \msolar black hole (the maximum mass considered
in this work) to approach the Eddington limit. To remove projection
effects, the black holes shown in Figure \ref{projection} are colored
by the distance to the nearest gas clump with accretability of at
least $10^{-3}$ M$^{-1}_{\odot}$. The closest encounter between a black
hole and a highly accretable clump is roughly 30 pc, but on average,
black holes are many hundreds of pc away from these clumps. In this
snapshot, no black holes are within such a clump.

In order for a black hole to experience high growth rates, it must
intersect with a high accretability clump at some time. 
To quantify the frequency of interactions between black holes and
clumps, we measure the distances from each black hole to the edge of
the nearest clump over the history of this halo and all of its
progenitors. We construct a merger-tree of this halo using the
\texttt{consistent-trees} merger-tree code
\citep{2013ApJ...763...18B}. We use the \texttt{ytree} code
\citep{ytree} to walk the tree, interface with \texttt{yt}, and run
the clump finding algorithm as described above for all progenitors of
the halo in question. In Figure \ref{fig:distances}, we plot the
distribution of distances between black holes and nearest clumps as a
function of time, with the median value for black holes within one
quarter of the virial radius shown in
black. Throughout most of the halo's history, black holes remain on
average a few hundred pc away from clumps in which they could grow
rapidly, with the closest black holes still tens of pc away. In total,
we note 10 occasions of black holes existing in highly accretable
clumps, with their details shown in Table \ref{tab:bhclump}. Half of
these ten occurrences consist of a new black hole in the direct
formation mass range. Two others in the same mass range are only 2.5
Myr old. The maximum time spent inside a clump was just over 3 Myr
with an average accretion rate of $10^{-8}$ \msolarc/yr, and in all
cases the accretion rate was sub-Eddington. Black holes appear to have
a difficult time remaining in clumps, either because they migrate out
or because those clumps are consumed or destroyed. If the latter is
true, then the most likely cause is star formation, as highly
accretable gas is cold and dense.

\begin{table*}
\caption{Black holes interacting with clumps in Figure
  \ref{fig:distances}. \label{tab:bhclump}}
\begin{tabular}{cccccccc}
\hline
Particle ID$^a$ & z$^b$ & m$_{\rm bh}^c$ [M$_{\odot}$] & age$^d$ [Myr] &
$\dot{\rm m}_{\rm bh}^e$
[M$_{\odot}$ / yr] & $\dot{\rm m}_{\rm bh}^f$ / $\dot{\rm m}_{\rm Edd}$
& $\Delta$t$^g$ [Myr] & $\Delta$m$^h$ [M$_{\odot}$]\\
\hline

618050081 & 15.2 & 46.053 & 0.00  & 1.167e-07 & 1.152e-01 & 2.547 & 2.973e-01\\
618051131 & 15.2 & 48.921 & 0.00  & 1.878e-07 & 1.745e-01 & 2.547 & 4.784e-01\\
618048385 & 15.2 & 28.510 & 0.00  & 8.338e-08 & 1.329e-01 & 2.547 & 2.124e-01\\
"         & 15.1 & 28.722 & 2.54  & 5.292e-08 & 8.374e-02 & 2.587 & 1.369e-01\\
618049168 & 15.2 & 27.457 & 0.00  & 5.915e-08 & 9.793e-02 & 2.547 & 1.507e-01\\
"         & 15.1 & 27.607 & 2.54  & 2.032e-08 & 3.346e-02 & 2.587 & 5.258e-02\\
618062820 & 15.1 & 17.272 & 0.00  & 4.052e-08 & 1.066e-01 & 2.587 & 1.048e-01\\
617970154 & 14.8 &  8.547 & 33.88 & 1.957e-09 & 1.041e-02 & 2.712 & 5.307e-03\\
617970240 & 13.8 & 42.303 & 67.36 & 5.976e-08 & 6.422e-02 & 3.195 & 1.910e-01\\
617977875 & 13.8 & 17.271 & 63.07 & 7.217e-09 & 1.899e-02 & 3.195 & 2.306e-02\\

\hline
\end{tabular}
\\ (a) the particle id of the black hole; (b) redshift of
interaction; (c) black hole mass; (d) black hole age; (e) Bondi-Hoyle
accretion rate; (f) fraction of Eddington accretion rate; (g)
time between current and next snapshots; (h) mass accreted in
$\Delta$t.
\end{table*}
\begin{figure*}
    \includegraphics[width=\textwidth]{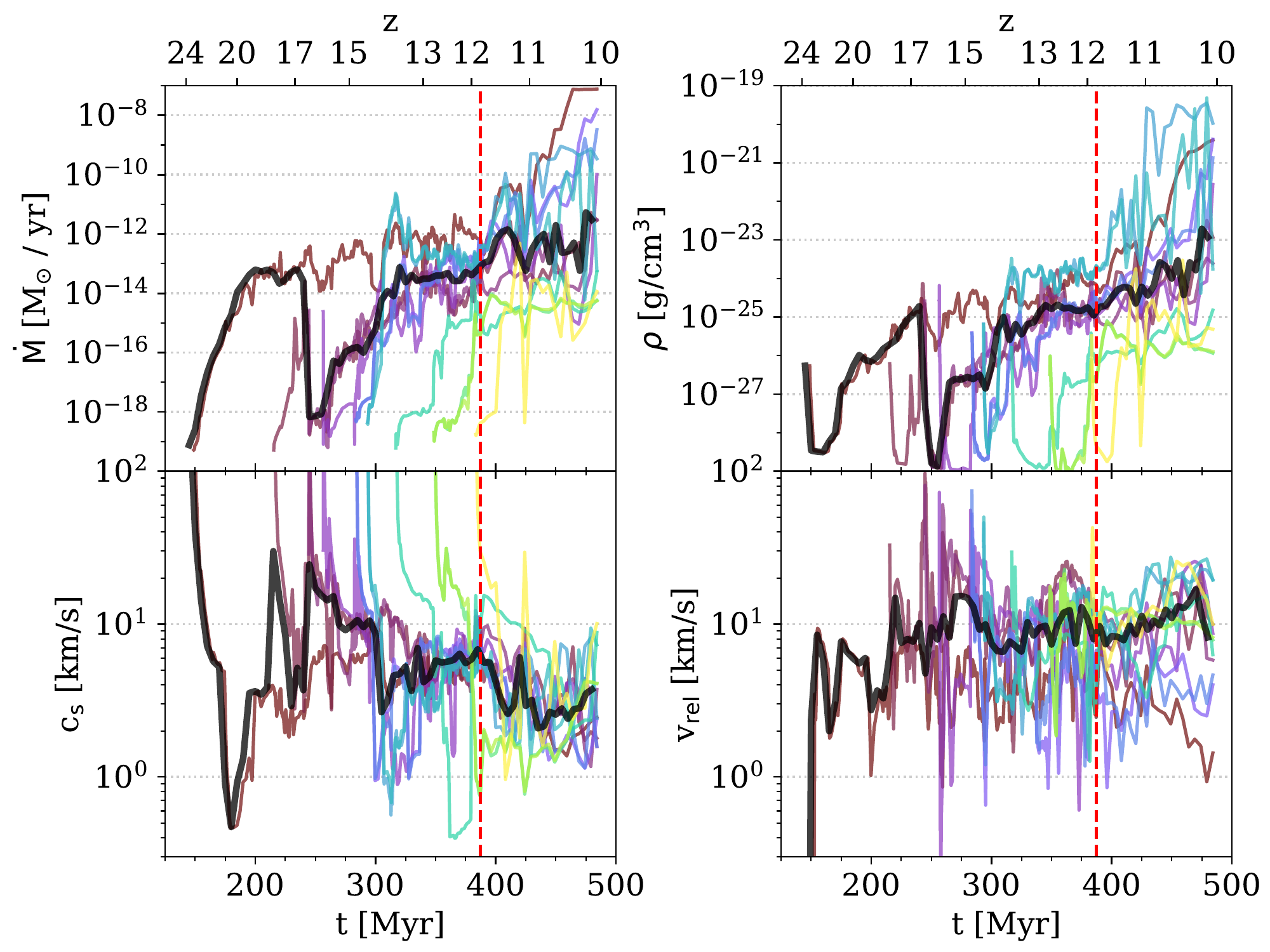}
    \caption[]
            {\label{p2p}
              Growth histories and associated gas physical conditions
              for all 12 black holes in the \pp simulation, with
              median values shown by the thick, black line. The
              vertical, red dashed line denotes the time when star
              formation was turned off. \textit{Top-left:} accretion
              rate. \textit{Top-right:} gas density.
              \textit{Bottom-left:} local sound
              speed. \textit{Bottom-right:} relative velocity between
              gas and particle.
             }
\end{figure*}

We use the \pp simulation to test the hypothesis that black hole
growth is regulate by star formation and feedback.  As described in \S
\ref{Sec:PopIIPrime}, the \pp simulation is an extremely high
resolution simulation in a very small volume and hence only 12 black
holes exist by $z \sim$ 12. At $z \simeq 11.8$, we turn off star
formation, but allow the simulation to evolve in every other
respect. The growth rates for all 12 black holes are shown in Figure
\ref{p2p}. After star formation is disabled, the mean growth rate
increases by roughly two orders of magnitude in 100 Myr. However, more
notably, the growth rates of the oldest black holes, whose haloes have
had much more time to reassemble, have increased by a much greater
degree. \citet{2014MNRAS.444.3288J} find that $\sim10^{5}$ \msolar
mini-haloes can take more than 100 Myr to reassemble, so the low
growth rates of the latest forming black holes are not surprising. The
top-right panel of Figure \ref{p2p} indicates that it is the increase
in density which drives the enhancement of black hole growth after
star formation has ceased. The sound speed has also dropped
considerably at this time, but this turns out to be unimportant as the
gas/particle relative velocity remains roughly 10 km/s. With a
self-consistent treatment of black hole formation and evolution (i.e.,
not what we have done here), black holes should eventually sink to the
center of the halo due to dynamical friction, although this will be slow
because of their low masses.
Indeed, \citet{2018MNRAS.478.3961S} find that dynamical friction is
unable to transport black holes formed in mini-halos into the inner,
gas-rich regions of atomic cooling halos.
As galaxies grow
larger, they will be more able to co-locate black holes and accretable
gas. Regardless, this provides strong evidence that the ability of
stellar feedback to destroy cold, dense gas through radiation and
supernovae is quite important in regulating black hole growth.

\section{Summary \& Discussion}  \label{Sec:Discussion}

The goal of this paper was to use the large sample of Pop III remnant
black holes provided by the \ren simulations to study both their
evolution and growth over the course in the early Universe. The \ren
simulations sample three large-scale galactic environments, span three
orders of magnitude in halo mass (10$^{6}$ \msolar $\le$ M$_{halo}$
$\le$ 10$^{9}$ \msolarc), provide roughly 300 Myr of black
hole evolution time, and form roughly 15,000 Pop III remnant black
holes. This mass range allows us to span the boundary between
\molHc-cooling mini-haloes and atomic cooling haloes. We supplement this
with 12 Pop III remnants from the extremely high-resolution \pp
simulation, in which we disable star formation part of the way through
to test the effect of stellar feedback on black hole growth.  Our
results can be summarized as:
\begin{itemize}

\item In all simulations, the vast majority of Pop III remnant black
  holes grow by negligible amounts, with relative mass gains ranging
  from 10$^{-16}$ to 0.1. Less than 100 of the 15,000 total black
  holes grew by more than 10$^{-3}$ of their initial masses, with most of
  these coming from the Rarepeak (densest region), followed by the
  Normal (next densest), and just a few from the Void simulation.
  The instantaneous accretion rates only exceed the Eddington rate one
  time for one black hole.

\item The black holes that grew the most did so within about 10
  Myr of their formation, after which time they grew negligibly. All
  formed from Pop III stars in the mass range where no supernova
  occurs. These black holes show no preference in host halo mass.

\item Clumps of gas with high Bondi-Hoyle accretion rates exist within
  galaxies, but the instances of black holes existing within them are
  rare and short-lived. On average, black holes are tens to hundreds
  of pc away from highly accretable (cold and dense) clumps. These
  clumps appear to be rapidly detroyed by star formation and feedback
  before black holes have a chance to accrete significantly from
  them. In examining the halo hosting the most black holes in the \ren
  simulations, most of the
  instances of black holes located within accretable clumps were newly
  born and formed without a supernova.

\item In the \pp simulations, the average black hole growth increased
  by more than two orders of magnitude within 100 Myr of turning off
  star formation.  Black holes in haloes that had more time to
  reassemble after the black hole-forming supernova showed
  significantly increased growth rates, up to five orders of
  magnitude.  This is due primarily to the increase in gas density.

\end{itemize}

Overall the \ren simulations indicate that black holes born from the
remnants of Pop III stars never enter regions where rapid and
sustained accretion is possible. The early bottleneck which has been
previously shown to prevent early black hole growth continues as black
holes migrate into more massive haloes, although the halos studied
here remain under the minimum mass where \citet{2007ApJ...665..107P}
find super-Eddington growth may be possible.
We conclude that black holes born from Pop III remnants in
mini-haloes are likely to experience very limited growth in the early
Universe. We see no appreciable growth in any black holes in halo
masses up to $10^9$ \msolarc. This provides further evidence of the
difficulty of Pop III remnant black holes reaching SMBH scales by $z =
6$, although the possibility of doing so in rarer, more massive haloes
still exists. This has additional consequences for the
formation of intermediate mass black holes (IMBHs). If the Pop III
remnants seeds are unable to accrete as cosmic time proceeds in more
massive haloes then they will not grow to form a population of IMBHs
and may go someway to explaining the relative dearth of IMBHs observed
thus far \citep{Koliopanos_2018}. It is also possible that over a
longer period these black holes will sink to the centre of larger
haloes or undergo several interactions with clumps which enable the
black hole to grow. However, \ren does not follow the evolution of
structure sufficiently far for us to answer that
question. Alternatively, it may also be that periods of high accretion
(possibly super-Eddington) can be achieved in metal-free haloes with
reduced \molH fractions rather than in the minihaloes that led to the
creation of the black holes examined here. However, \ren has no
prescription to form stars in such environments, though the
environments do exist in \ren (Wise et al. in prep). Such regions with
their deeper potential wells may be more advantageous for forming
rapidly accreting black hole seeds which can then go on to become IMBH
and/or SMBHs.

A number of other simplifications made in this work or by the
simulations studied have the potential to alter the findings presented
here. The use of post-processing limits the temporal resolution to the
time between snapshots, which here is generally a few Myr. Any events
occuring between snaphots will naturally be lost. In particular, this
likely leads to an underestimation of the early-time growth of black
holes forming without a supernova, i.e., the situation that we find to
yield the most growth.
As well, any other dynamic effects requiring an accurate calculation
of the black hole's mass evolution during the simulation, such as
dynamical friction, cannot be accounted for.
The influence of feedback from the accreting black hole
on the survival of high accretability clouds is also not clear. Black
hole feedback could act similarly to stellar feedback and destroy
these clouds, or it may simply delay star formation while largely
keeping the cloud in tact. This could potentially give any embedded
black holes additional time to grow. Finally, the optically-thin
treatment of the LW radiation field may lead to artificially low \molH
fractions in some instances, thus reducing the cooling rate in
$\sim10^{3}$ K gas and lowering both the star formation and black hole
growth efficiencies. A significantly higher \molH fraction could
potentially even give rise to thermal instability, which could require
the use of a boost factor higher than one for the accretion rate.
However, an increase in the cooling would also allow stars to
form more readily, which will act to destroy high-accretability gas,
making the precise balance unclear. As always, these and any other
shortcomings provide ample avenues for further progress.

\section*{Acknowledgements}
We thank the anonymous referee for their useful comments on the
manuscript.
BDS is supported by NSF AST-1615848. JAR acknowledges the support of
the EU Commission through the Marie Sk\l{}odowska-Curie Grant -
``SMARTSTARS" - grant number 699941. MLN was supported by NSF
AST-1102943 and AST-1615848. BWO was supported in part by NSF grants
PHY-1430152 and AST-1514700, by NASA grants NNX12AC98G, NNX15AP39G,
and by Hubble Theory Grants HST-AR-13261.01-A and
HST-AR-14315.001-A. JHW is supported by National Science Foundation
grant AST-1614333, NASA grant NNX17AG23G, and Hubble theory grants
HST-AR-13895 and HST-AR-14326.
The \ren and \pp simulations were performed on Blue Waters,
operated by the National Center for Supercomputing Applications
(NCSA) with PRAC allocation support by the NSF (award number
ACI-0832662). The subsequent analysis was performed using time awarded
under award number ACI-1514580. This research is part of the Blue
Waters sustained-petascale computing project, which is supported by
the NSF (award number ACI-1238993 and ACI-1514580) and the state of
Illinois. Blue Waters is a joint effort of the University of
Illinois at Urbana-Champaign and its NCSA.
Finally, this research was made possible by a mountain of open-source
scientific software, including \enzo \citep{Enzo_2014}, \texttt{yt}
\citep{YT}, \texttt{Rockstar} \citep{2013ApJ...762..109B},
\texttt{consistent-trees} \citep{2013ApJ...763...18B}, and
\texttt{ytree} \citep{ytree}, sitting on a vast continent of
open-source software packages and libraries like \texttt{Python},
\texttt{matplotlib}, and \texttt{NumPy}, just to name a few.  We are
grateful to everyone who contributed to these projects.


\noindent

\begin{thebibliography}{114}
\providecommand{\natexlab}[1]{#1}

\bibitem[{{Abel} et~al.(1997){Abel}, {Anninos}, {Zhang} \&
  {Norman}}]{Abel_1997}
{Abel} T., {Anninos} P., {Zhang} Y., {Norman} M.~L., 1997, New Astronomy, 2,
  181

\bibitem[{{Abel} et~al.(2000){Abel}, {Bryan} \& {Norman}}]{Abel_2000}
{Abel} T., {Bryan} G.~L., {Norman} M.~L., 2000, \apj, 540, 39

\bibitem[{{Abel} et~al.(2002{\natexlab{a}}){Abel}, {Bryan} \&
  {Norman}}]{Abel_2002}
{Abel} T., {Bryan} G.~L., {Norman} M.~L., 2002{\natexlab{a}}, Science, 295, 93

\bibitem[{{Abel} et~al.(2002{\natexlab{b}}){Abel}, {Bryan} \&
  {Norman}}]{2002Sci...295...93A}
{Abel} T., {Bryan} G.~L., {Norman} M.~L., 2002{\natexlab{b}}, Science, 295, 93

\bibitem[{{Agarwal} et~al.(2012){Agarwal}, {Khochfar}, {Johnson}, {Neistein},
  {Dalla Vecchia} \& {Livio}}]{Agarwal_2012}
{Agarwal} B., {Khochfar} S., {Johnson} J.~L., {Neistein} E., {Dalla Vecchia}
  C., {Livio} M., 2012, \mnras, 425, 2854

\bibitem[{{Agarwal} et~al.(2013){Agarwal}, {Davis}, {Khochfar}, {Natarajan} \&
  {Dunlop}}]{Agarwal_2013}
{Agarwal} B., {Davis} A.~J., {Khochfar} S., {Natarajan} P., {Dunlop} J.~S.,
  2013, \mnras, 432, 3438

\bibitem[{{Agarwal} et~al.(2016){Agarwal}, {Smith}, {Glover}, {Natarajan} \&
  {Khochfar}}]{Agarwal_2015b}
{Agarwal} B., {Smith} B., {Glover} S., {Natarajan} P., {Khochfar} S., 2016,
  \mnras, 459, 4209

\bibitem[{{Ahn} et~al.(2015){Ahn}, {Xu}, {Norman}, {Alvarez} \&
  {Wise}}]{Ahn_2015}
{Ahn} K., {Xu} H., {Norman} M.~L., {Alvarez} M.~A., {Wise} J.~H., 2015, \apj,
  802, 8

\bibitem[{{Alvarez} et~al.(2009){Alvarez}, {Wise} \& {Abel}}]{Alvarez_2009}
{Alvarez} M.~A., {Wise} J.~H., {Abel} T., 2009, \apjl, 701, L133

\bibitem[{{Ba{\~n}ados} et~al.(2018)}]{Banados_2018}
{Ba{\~n}ados} E. et~al., 2018, \nat, 553, 473

\bibitem[{{Becerra} et~al.(2015){Becerra}, {Greif}, {Springel} \&
  {Hernquist}}]{Becerra_2015}
{Becerra} F., {Greif} T.~H., {Springel} V., {Hernquist} L.~E., 2015, \mnras,
  446, 2380

\bibitem[{{Begelman} et~al.(2006){Begelman}, {Volonteri} \&
  {Rees}}]{Begelman_2006}
{Begelman} M.~C., {Volonteri} M., {Rees} M.~J., 2006, \mnras, 370, 289

\bibitem[{{Begelman} et~al.(2008){Begelman}, {Rossi} \&
  {Armitage}}]{Begelman_2008}
{Begelman} M.~C., {Rossi} E.~M., {Armitage} P.~J., 2008, \mnras, 387, 1649

\bibitem[{{Behroozi} et~al.(2013{\natexlab{a}}){Behroozi}, {Wechsler} \&
  {Wu}}]{2013ApJ...762..109B}
{Behroozi} P.~S., {Wechsler} R.~H., {Wu} H.~Y., 2013{\natexlab{a}}, \apj, 762,
  109

\bibitem[{{Behroozi} et~al.(2013{\natexlab{b}}){Behroozi}, {Wechsler}, {Wu},
  {Busha}, {Klypin} \& {Primack}}]{2013ApJ...763...18B}
{Behroozi} P.~S., {Wechsler} R.~H., {Wu} H.~Y., {Busha} M.~T., {Klypin} A.~A.,
  {Primack} J.~R., 2013{\natexlab{b}}, \apj, 763, 18

\bibitem[{{Bondi}(1952)}]{1952MNRAS.112..195B}
{Bondi} H., 1952, \mnras, 112, 195

\bibitem[{{Booth} \& {Schaye}(2009)}]{2009MNRAS.398...53B}
{Booth} C.~M., {Schaye} J., 2009, \mnras, 398, 53

\bibitem[{{Bromm} et~al.(1999){Bromm}, {Coppi} \& {Larson}}]{Bromm_1999}
{Bromm} V., {Coppi} P.~S., {Larson} R.~B., 1999, \apjl, 527, L5

\bibitem[{{Bromm} et~al.(2002){Bromm}, {Coppi} \& {Larson}}]{Bromm_2002}
{Bromm} V., {Coppi} P.~S., {Larson} R.~B., 2002, \apj, 564, 23

\bibitem[{{Bryan} et~al.(2014){Bryan}, {Norman}, {O'Shea}, {Abel}, {Wise},
  {Turk} \& {The Enzo Collaboration}}]{Enzo_2014}
{Bryan} G.~L., {Norman} M.~L., {O'Shea} B.~W., {Abel} T., {Wise} J.~H., {Turk}
  M.~J., {The Enzo Collaboration}, 2014, \apjs, 211, 19

\bibitem[{{Chen} et~al.(2014{\natexlab{a}}){Chen}, {Wise}, {Norman}, {Xu} \&
  {O'Shea}}]{2014ApJ...795..144C}
{Chen} P., {Wise} J.~H., {Norman} M.~L., {Xu} H., {O'Shea} B.~W.,
  2014{\natexlab{a}}, \apj, 795, 144

\bibitem[{{Chen} et~al.(2014{\natexlab{b}}){Chen}, {Wise}, {Norman}, {Xu} \&
  {O'Shea}}]{Chen_2014}
{Chen} P., {Wise} J.~H., {Norman} M.~L., {Xu} H., {O'Shea} B.~W.,
  2014{\natexlab{b}}, \apj, 795, 144

\bibitem[{{Clark} et~al.(2011){Clark}, {Glover}, {Smith}, {Greif}, {Klessen} \&
  {Bromm}}]{Clark_2011a}
{Clark} P.~C., {Glover} S.~C.~O., {Smith} R.~J., {Greif} T.~H., {Klessen}
  R.~S., {Bromm} V., 2011, Science, 331, 1040

\bibitem[{{Crosby} et~al.(2013){Crosby}, {O'Shea}, {Smith}, {Turk} \&
  {Hahn}}]{2013ApJ...773..108C}
{Crosby} B.~D., {O'Shea} B.~W., {Smith} B.~D., {Turk} M.~J., {Hahn} O., 2013,
  \apj, 773, 108

\bibitem[{{Devecchi} \& {Volonteri}(2009)}]{Devecchi_2008}
{Devecchi} B., {Volonteri} M., 2009, \apj, 694, 302

\bibitem[{{Eisenstein} \& {Hu}(1999)}]{1999ApJ...511....5E}
{Eisenstein} D.~J., {Hu} W., 1999, \apj, 511, 5

\bibitem[{{Fan} et~al.(2006){Fan}, {Carilli} \& {Keating}}]{Fan_2006}
{Fan} X., {Carilli} C.~L., {Keating} B., 2006, \araa, 44, 415

\bibitem[{{Fielding} et~al.(2017){Fielding}, {Quataert}, {McCourt} \&
  {Thompson}}]{2017MNRAS.466.3810F}
{Fielding} D., {Quataert} E., {McCourt} M., {Thompson} T.~A., 2017, \mnras,
  466, 3810

\bibitem[{{G{\"u}rkan} et~al.(2004){G{\"u}rkan}, {Freitag} \&
  {Rasio}}]{Gurkan_2004}
{G{\"u}rkan} M.~A., {Freitag} M., {Rasio} F.~A., 2004, \apj, 604, 632

\bibitem[{{G{\"u}rkan} et~al.(2006){G{\"u}rkan}, {Fregeau} \&
  {Rasio}}]{Gurkan_2006}
{G{\"u}rkan} M.~A., {Fregeau} J.~M., {Rasio} F.~A., 2006, \apjl, 640, L39

\bibitem[{{Habouzit} et~al.(2017){Habouzit}, {Volonteri} \&
  {Dubois}}]{2017MNRAS.468.3935H}
{Habouzit} M., {Volonteri} M., {Dubois} Y., 2017, \mnras, 468, 3935

\bibitem[{{Haemmerl{\'e}} et~al.(2018){Haemmerl{\'e}}, {Woods}, {Klessen},
  {Heger} \& {Whalen}}]{Haemmerle_2017}
{Haemmerl{\'e}} L., {Woods} T.~E., {Klessen} R.~S., {Heger} A., {Whalen} D.~J.,
  2018, \mnras, 474, 2757

\bibitem[{{Hahn} \& {Abel}(2011)}]{Hahn_2011}
{Hahn} O., {Abel} T., 2011, \mnras, 415, 2101

\bibitem[{{Heger} \& {Woosley}(2002)}]{2002ApJ...567..532H}
{Heger} A., {Woosley} S.~E., 2002, \apj, 567, 532

\bibitem[{{Hirano} et~al.(2014){Hirano}, {Hosokawa}, {Yoshida}, {Umeda},
  {Omukai}, {Chiaki} \& {Yorke}}]{Hirano_2014}
{Hirano} S., {Hosokawa} T., {Yoshida} N., {Umeda} H., {Omukai} K., {Chiaki} G.,
  {Yorke} H.~W., 2014, \apj, 781, 60

\bibitem[{{Hosokawa} et~al.(2013{\natexlab{a}}){Hosokawa}, {Omukai} \&
  {Yorke}}]{Hosokawa_2012}
{Hosokawa} T., {Omukai} K., {Yorke} H.~W., 2013{\natexlab{a}}, \apj, 778, 178

\bibitem[{{Hosokawa} et~al.(2013{\natexlab{b}}){Hosokawa}, {Yorke}, {Inayoshi},
  {Omukai} \& {Yoshida}}]{Hosokawa_2013}
{Hosokawa} T., {Yorke} H.~W., {Inayoshi} K., {Omukai} K., {Yoshida} N.,
  2013{\natexlab{b}}, \apj, 778, 178

\bibitem[{{Hosokawa} et~al.(2016){Hosokawa}, {Hirano}, {Kuiper}, {Yorke},
  {Omukai} \& {Yoshida}}]{Hosokawa_2015}
{Hosokawa} T., {Hirano} S., {Kuiper} R., {Yorke} H.~W., {Omukai} K., {Yoshida}
  N., 2016, \apj, 824, 119

\bibitem[{{Hoyle} \& {Lyttleton}(1941)}]{1941MNRAS.101..227H}
{Hoyle} F., {Lyttleton} R.~A., 1941, \mnras, 101, 227

\bibitem[{{Jeon} et~al.(2012){Jeon}, {Pawlik}, {Greif}, {Glover}, {Bromm},
  {Milosavljevi{\'c}} \& {Klessen}}]{2012ApJ...754...34J}
{Jeon} M., {Pawlik} A.~H., {Greif} T.~H., {Glover} S.~C.~O., {Bromm} V.,
  {Milosavljevi{\'c}} M., {Klessen} R.~S., 2012, \apj, 754, 34

\bibitem[{{Jeon} et~al.(2014){Jeon}, {Pawlik}, {Bromm} \&
  {Milosavljevi{\'c}}}]{2014MNRAS.444.3288J}
{Jeon} M., {Pawlik} A.~H., {Bromm} V., {Milosavljevi{\'c}} M., 2014, \mnras,
  444, 3288

\bibitem[{{Johnson} \& {Bromm}(2007)}]{Johnson_2007}
{Johnson} J.~L., {Bromm} V., 2007, \mnras, 374, 1557

\bibitem[{{Katz} et~al.(2015){Katz}, {Sijacki} \& {Haehnelt}}]{Katz_2015}
{Katz} H., {Sijacki} D., {Haehnelt} M.~G., 2015, \mnras, 451, 2352

\bibitem[{{Kim} et~al.(2011){Kim}, {Wise}, {Alvarez} \&
  {Abel}}]{2011ApJ...738...54K}
{Kim} J.~h., {Wise} J.~H., {Alvarez} M.~A., {Abel} T., 2011, \apj, 738, 54

\bibitem[{{Koliopanos}(2018)}]{Koliopanos_2018}
{Koliopanos} F., 2018, ArXiv e-prints

\bibitem[{{Komatsu} et~al.(2011{\natexlab{a}}){Komatsu}, {Smith}, {Dunkley},
  {Bennett}, {Gold} \& {Hinshaw}}]{2011ApJS..192...18K}
{Komatsu} E., {Smith} K.~M., {Dunkley} J., {Bennett} C.~L., {Gold} B.,
  {Hinshaw}, 2011{\natexlab{a}}, \apjs, 192, 18

\bibitem[{{Komatsu} et~al.(2011{\natexlab{b}})}]{Komatsu_2011}
{Komatsu} E. et~al., 2011{\natexlab{b}}, \apjs, 192, 18

\bibitem[{{Krumholz} et~al.(2005){Krumholz}, {McKee} \&
  {Klein}}]{Krumholz_2005}
{Krumholz} M.~R., {McKee} C.~F., {Klein} R.~I., 2005, \apj, 618, 757

\bibitem[{{Latif} et~al.(2013{\natexlab{a}}){Latif}, {Schleicher}, {Schmidt} \&
  {Niemeyer}}]{Latif_2013c}
{Latif} M.~A., {Schleicher} D.~R.~G., {Schmidt} W., {Niemeyer} J.,
  2013{\natexlab{a}}, \mnras, 433, 1607

\bibitem[{{Latif} et~al.(2013{\natexlab{b}}){Latif}, {Schleicher}, {Schmidt} \&
  {Niemeyer}}]{Latif_2013a}
{Latif} M.~A., {Schleicher} D.~R.~G., {Schmidt} W., {Niemeyer} J.,
  2013{\natexlab{b}}, \mnras, 430, 588

\bibitem[{{Lupi} et~al.(2016){Lupi}, {Haardt}, {Dotti}, {Fiacconi}, {Mayer} \&
  {Madau}}]{Lupi_2016}
{Lupi} A., {Haardt} F., {Dotti} M., {Fiacconi} D., {Mayer} L., {Madau} P.,
  2016, \mnras, 456, 2993

\bibitem[{{Machacek} et~al.(2001){Machacek}, {Bryan} \& {Abel}}]{Machacek_2001}
{Machacek} M.~E., {Bryan} G.~L., {Abel} T., 2001, \apj, 548, 509

\bibitem[{{Madau} \& {Rees}(2001)}]{Madau_2001}
{Madau} P., {Rees} M.~J., 2001, \apjl, 551, L27

\bibitem[{{McCourt} et~al.(2018){McCourt}, {Oh}, {O'Leary} \&
  {Madigan}}]{2018MNRAS.473.5407M}
{McCourt} M., {Oh} S.~P., {O'Leary} R., {Madigan} A.~M., 2018, \mnras, 473,
  5407

\bibitem[{{Meece} et~al.(2014){Meece}, {Smith} \&
  {O'Shea}}]{2014ApJ...783...75M}
{Meece} G.~R., {Smith} B.~D., {O'Shea} B.~W., 2014, \apj, 783, 75

\bibitem[{{Milosavljevi{\'c}} et~al.(2009){Milosavljevi{\'c}}, {Couch} \&
  {Bromm}}]{Milosavljevic_2009}
{Milosavljevi{\'c}} M., {Couch} S.~M., {Bromm} V., 2009, \apjl, 696, L146

\bibitem[{{Mortlock} et~al.(2011)}]{Mortlock_2011}
{Mortlock} D.~J. et~al., 2011, \nat, 474, 616

\bibitem[{{Nakauchi} et~al.(2017){Nakauchi}, {Hosokawa}, {Omukai}, {Saio} \&
  {Nomoto}}]{Nakauchi_2017}
{Nakauchi} D., {Hosokawa} T., {Omukai} K., {Saio} H., {Nomoto} K., 2017,
  \mnras, 465, 5016

\bibitem[{{Nomoto} et~al.(2006){Nomoto}, {Tominaga}, {Umeda}, {Kobayashi} \&
  {Maeda}}]{2006NuPhA.777..424N}
{Nomoto} K., {Tominaga} N., {Umeda} H., {Kobayashi} C., {Maeda} K., 2006,
  Nuclear Physics A, 777, 424

\bibitem[{{O'Shea} \& {Norman}(2007{\natexlab{a}})}]{OShea_2007b}
{O'Shea} B.~W., {Norman} M.~L., 2007{\natexlab{a}}, \apj, 654, 66

\bibitem[{{O'Shea} \& {Norman}(2007{\natexlab{b}})}]{2007ApJ...654...66O}
{O'Shea} B.~W., {Norman} M.~L., 2007{\natexlab{b}}, \apj, 654, 66

\bibitem[{{O'Shea} \& {Norman}(2008{\natexlab{a}})}]{2008ApJ...673...14O}
{O'Shea} B.~W., {Norman} M.~L., 2008{\natexlab{a}}, \apj, 673, 14

\bibitem[{{O'Shea} \& {Norman}(2008{\natexlab{b}})}]{OShea_2008}
{O'Shea} B.~W., {Norman} M.~L., 2008{\natexlab{b}}, \apj, 673, 14

\bibitem[{{O'Shea} et~al.(2005){O'Shea}, {Abel}, {Whalen} \&
  {Norman}}]{2005ApJ...628L...5O}
{O'Shea} B.~W., {Abel} T., {Whalen} D., {Norman} M.~L., 2005, \apjl, 628, L5

\bibitem[{{O'Shea} et~al.(2015){O'Shea}, {Wise}, {Xu} \& {Norman}}]{OShea_2015}
{O'Shea} B.~W., {Wise} J.~H., {Xu} H., {Norman} M.~L., 2015, \apjl, 807, L12

\bibitem[{{Pacucci} et~al.(2017){Pacucci}, {Natarajan}, {Volonteri},
  {Cappelluti} \& {Urry}}]{Pacucci_2017}
{Pacucci} F., {Natarajan} P., {Volonteri} M., {Cappelluti} N., {Urry} C.~M.,
  2017, \apjl, 850, L42

\bibitem[{{Pelupessy} et~al.(2007){Pelupessy}, {Di Matteo} \&
  {Ciardi}}]{2007ApJ...665..107P}
{Pelupessy} F.~I., {Di Matteo} T., {Ciardi} B., 2007, \apj, 665, 107

\bibitem[{{Pezzulli} et~al.(2016){Pezzulli}, {Valiante} \&
  {Schneider}}]{Pezzulli_2016}
{Pezzulli} E., {Valiante} R., {Schneider} R., 2016, \mnras, 458, 3047

\bibitem[{{Pezzulli} et~al.(2017){Pezzulli}, {Volonteri}, {Schneider} \&
  {Valiante}}]{Pezzulli_2017}
{Pezzulli} E., {Volonteri} M., {Schneider} R., {Valiante} R., 2017, \mnras,
  471, 589

\bibitem[{{Regan} \& {Haehnelt}(2009{\natexlab{a}})}]{Regan_2009b}
{Regan} J.~A., {Haehnelt} M.~G., 2009{\natexlab{a}}, \mnras, 396, 343

\bibitem[{{Regan} \& {Haehnelt}(2009{\natexlab{b}})}]{Regan_2009}
{Regan} J.~A., {Haehnelt} M.~G., 2009{\natexlab{b}}, \mnras, 393, 858

\bibitem[{{Regan} et~al.(2014){Regan}, {Johansson} \& {Haehnelt}}]{Regan_2014a}
{Regan} J.~A., {Johansson} P.~H., {Haehnelt} M.~G., 2014, \mnras, 439, 1160

\bibitem[{{Regan} et~al.(2016){Regan}, {Johansson} \& {Wise}}]{Regan_2016a}
{Regan} J.~A., {Johansson} P.~H., {Wise} J.~H., 2016, \mnras

\bibitem[{{Regan} et~al.(2017){Regan}, {Visbal}, {Wise}, , {Haiman},
  {Johansson} \& {Bryan}}]{Regan_2017}
{Regan} J.~A., {Visbal} E., {Wise} J.~H., , {Haiman} Z., {Johansson} P.~H.,
  {Bryan} G.~L., 2017, Nature Astronomy, 1, 0075

\bibitem[{{Schleicher} et~al.(2013){Schleicher}, {Palla}, {Ferrara}, {Galli} \&
  {Latif}}]{Schleicher_2013}
{Schleicher} D.~R.~G., {Palla} F., {Ferrara} A., {Galli} D., {Latif} M., 2013,
  \aap, 558, A59

\bibitem[{{Shakura} \& {Syunyaev}(1973)}]{Shakura_1973}
{Shakura} N.~I., {Syunyaev} R.~A., 1973, \aap, 24, 337

\bibitem[{Smith(2018)}]{ytree}
Smith B., 2018, https://github.com/brittonsmith/ytree: ytree version 2.0.2
  release

\bibitem[{{Smith} \& {Sigurdsson}(2007)}]{2007ApJ...661L...5S}
{Smith} B.~D., {Sigurdsson} S., 2007, \apjl, 661, L5

\bibitem[{{Smith} et~al.(2009{\natexlab{a}}){Smith}, {Turk}, {Sigurdsson},
  {O'Shea} \& {Norman}}]{2009ApJ...691..441S}
{Smith} B.~D., {Turk} M.~J., {Sigurdsson} S., {O'Shea} B.~W., {Norman} M.~L.,
  2009{\natexlab{a}}, \apj, 691, 441

\bibitem[{{Smith} et~al.(2009{\natexlab{b}}){Smith}, {Turk}, {Sigurdsson},
  {O'Shea} \& {Norman}}]{Smith_2009}
{Smith} B.~D., {Turk} M.~J., {Sigurdsson} S., {O'Shea} B.~W., {Norman} M.~L.,
  2009{\natexlab{b}}, \apj, 691, 441

\bibitem[{{Smith} et~al.(2015){Smith}, {Wise}, {O'Shea}, {Norman} \&
  {Khochfar}}]{Smith_2015}
{Smith} B.~D., {Wise} J.~H., {O'Shea} B.~W., {Norman} M.~L., {Khochfar} S.,
  2015, \mnras, 452, 2822

\bibitem[{{Springel} et~al.(2005){Springel}, {Di Matteo} \&
  {Hernquist}}]{2005MNRAS.361..776S}
{Springel} V., {Di Matteo} T., {Hernquist} L., 2005, \mnras, 361, 776

\bibitem[{{Sugimura} et~al.(2018){Sugimura}, {Hosokawa}, {Yajima}, {Inayoshi}
  \& {Omukai}}]{2018MNRAS.478.3961S}
{Sugimura} K., {Hosokawa} T., {Yajima} H., {Inayoshi} K., {Omukai} K., 2018,
  \mnras, 478, 3961

\bibitem[{{Tanaka} \& {Haiman}(2009)}]{Tanaka_2008}
{Tanaka} T., {Haiman} Z., 2009, \apj, 696, 1798

\bibitem[{{Tegmark} et~al.(1997){Tegmark}, {Silk}, {Rees}, {Blanchard}, {Abel}
  \& {Palla}}]{Tegmark_1997}
{Tegmark} M., {Silk} J., {Rees} M.~J., {Blanchard} A., {Abel} T., {Palla} F.,
  1997, \apj, 474, 1

\bibitem[{{Trenti} \& {Stiavelli}(2009)}]{2009ApJ...694..879T}
{Trenti} M., {Stiavelli} M., 2009, \apj, 694, 879

\bibitem[{{Turk} et~al.(2009{\natexlab{a}}){Turk}, {Abel} \&
  {O'Shea}}]{Turk_2009}
{Turk} M.~J., {Abel} T., {O'Shea} B., 2009{\natexlab{a}}, Science, 325, 601

\bibitem[{{Turk} et~al.(2009{\natexlab{b}}){Turk}, {Abel} \&
  {O'Shea}}]{2009Sci...325..601T}
{Turk} M.~J., {Abel} T., {O'Shea} B., 2009{\natexlab{b}}, Science, 325, 601

\bibitem[{{Turk} et~al.(2010){Turk}, {Norman} \& {Abel}}]{2010ApJ...725L.140T}
{Turk} M.~J., {Norman} M.~L., {Abel} T., 2010, \apjl, 725, L140

\bibitem[{{Turk} et~al.(2011{\natexlab{a}}){Turk}, {Clark}, {Glover}, {Greif},
  {Abel}, {Klessen} \& {Bromm}}]{2011ApJ...726...55T}
{Turk} M.~J., {Clark} P., {Glover} S.~C.~O., {Greif} T.~H., {Abel} T.,
  {Klessen} R., {Bromm} V., 2011{\natexlab{a}}, \apj, 726, 55

\bibitem[{{Turk} et~al.(2011{\natexlab{b}}){Turk}, {Smith}, {Oishi}, {Skory},
  {Skillman}, {Abel} \& {Norman}}]{YT}
{Turk} M.~J., {Smith} B.~D., {Oishi} J.~S., {Skory} S., {Skillman} S.~W.,
  {Abel} T., {Norman} M.~L., 2011{\natexlab{b}}, \apjs, 192, 9

\bibitem[{{Turk} et~al.(2012){Turk}, {Oishi}, {Abel} \&
  {Bryan}}]{2012ApJ...745..154T}
{Turk} M.~J., {Oishi} J.~S., {Abel} T., {Bryan} G.~L., 2012, \apj, 745, 154

\bibitem[{{Valiante} et~al.(2016){Valiante}, {Schneider}, {Volonteri} \&
  {Omukai}}]{Valiante_2016}
{Valiante} R., {Schneider} R., {Volonteri} M., {Omukai} K., 2016, \mnras

\bibitem[{{Venemans} et~al.(2013)}]{Venemans_2013}
{Venemans} B.~P. et~al., 2013, \apj, 779, 24

\bibitem[{{Voit} et~al.(2017){Voit}, {Meece}, {Li}, {O'Shea}, {Bryan} \&
  {Donahue}}]{2017ApJ...845...80V}
{Voit} G.~M., {Meece} G., {Li} Y., {O'Shea} B.~W., {Bryan} G.~L., {Donahue} M.,
  2017, \apj, 845, 80

\bibitem[{{Volonteri} et~al.(2015){Volonteri}, {Silk} \&
  {Dubus}}]{Volonteri_2015}
{Volonteri} M., {Silk} J., {Dubus} G., 2015, \apj, 804, 148

\bibitem[{{Whalen} et~al.(2004){Whalen}, {Abel} \& {Norman}}]{Whalen_2004}
{Whalen} D., {Abel} T., {Norman} M.~L., 2004, \apj, 610, 14

\bibitem[{{Wise} \& {Abel}(2007)}]{Wise_2007b}
{Wise} J.~H., {Abel} T., 2007, \apj, 671, 1559

\bibitem[{{Wise} \& {Abel}(2011)}]{WiseAbel_2011}
{Wise} J.~H., {Abel} T., 2011, \mnras, 414, 3458

\bibitem[{{Wise} et~al.(2008){Wise}, {Turk} \& {Abel}}]{Wise_2008a}
{Wise} J.~H., {Turk} M.~J., {Abel} T., 2008, \apj, 682, 745

\bibitem[{Wise et~al.(2012)Wise, Abel, Turk, Norman \&
  Smith}]{2012MNRAS.427..311W}
Wise J.~H., Abel T., Turk M.~J., Norman M.~L., Smith B.~D., 2012, MNRAS, 427,
  311

\bibitem[{{Wise} et~al.(2012{\natexlab{a}}){Wise}, {Turk}, {Norman} \&
  {Abel}}]{2012ApJ...745...50W}
{Wise} J.~H., {Turk} M.~J., {Norman} M.~L., {Abel} T., 2012{\natexlab{a}},
  \apj, 745, 50

\bibitem[{{Wise} et~al.(2012{\natexlab{b}}){Wise}, {Turk}, {Norman} \&
  {Abel}}]{Wise_2012b}
{Wise} J.~H., {Turk} M.~J., {Norman} M.~L., {Abel} T., 2012{\natexlab{b}},
  \apj, 745, 50

\bibitem[{{Wise} et~al.(2014){Wise}, {Demchenko}, {Halicek}, {Norman}, {Turk},
  {Abel} \& {Smith}}]{2014MNRAS.442.2560W}
{Wise} J.~H., {Demchenko} V.~G., {Halicek} M.~T., {Norman} M.~L., {Turk} M.~J.,
  {Abel} T., {Smith} B.~D., 2014, \mnras, 442, 2560

\bibitem[{{Wolcott-Green} et~al.(2011){Wolcott-Green}, {Haiman} \&
  {Bryan}}]{Wolcott-Green_2011}
{Wolcott-Green} J., {Haiman} Z., {Bryan} G.~L., 2011, \mnras, 418, 838

\bibitem[{{Woods} et~al.(2017){Woods}, {Heger}, {Whalen}, {Haemmerl{\'e}} \&
  {Klessen}}]{Woods_2017}
{Woods} T.~E., {Heger} A., {Whalen} D.~J., {Haemmerl{\'e}} L., {Klessen} R.~S.,
  2017, \apjl, 842, L6

\bibitem[{{Woosley} \& {Weaver}(1995)}]{1995ApJS..101..181W}
{Woosley} S.~E., {Weaver} T.~A., 1995, \apjs, 101, 181

\bibitem[{{Wu} et~al.(2015)}]{Wu_2015}
{Wu} X.~B. et~al., 2015, \nat, 518, 512

\bibitem[{{Xu} et~al.(2013{\natexlab{a}}){Xu}, {Wise} \& {Norman}}]{Xu_2013}
{Xu} H., {Wise} J.~H., {Norman} M.~L., 2013{\natexlab{a}}, \apj, 773, 83

\bibitem[{{Xu} et~al.(2013{\natexlab{b}}){Xu}, {Wise} \&
  {Norman}}]{2013ApJ...773...83X}
{Xu} H., {Wise} J.~H., {Norman} M.~L., 2013{\natexlab{b}}, \apj, 773, 83

\bibitem[{{Xu} et~al.(2014){Xu}, {Ahn}, {Wise}, {Norman} \& {O'Shea}}]{Xu_2014}
{Xu} H., {Ahn} K., {Wise} J.~H., {Norman} M.~L., {O'Shea} B.~W., 2014, \apj,
  791, 110

\bibitem[{{Xu} et~al.(2016{\natexlab{a}}){Xu}, {Norman}, {O'Shea} \&
  {Wise}}]{Xu_2016b}
{Xu} H., {Norman} M.~L., {O'Shea} B.~W., {Wise} J.~H., 2016{\natexlab{a}},
  \apj, 823, 140

\bibitem[{{Xu} et~al.(2016{\natexlab{b}}){Xu}, {Wise}, {Norman}, {Ahn} \&
  {O'Shea}}]{Xu_2016}
{Xu} H., {Wise} J.~H., {Norman} M.~L., {Ahn} K., {O'Shea} B.~W.,
  2016{\natexlab{b}}, \apj, 833, 84

\bibitem[{{Yoshida} et~al.(2003){Yoshida}, {Abel}, {Hernquist} \&
  {Sugiyama}}]{Yoshida_2003a}
{Yoshida} N., {Abel} T., {Hernquist} L., {Sugiyama} N., 2003, \apj, 592, 645

\end{thebibliography}

\label{lastpage}
\end{document}